\documentclass[twocolumn,prb,showpacs]{revtex4}
\usepackage{graphicx}
\usepackage{dcolumn}
\usepackage{bm}
\usepackage{amsmath}

\def\simge{\lower0.7ex\hbox{$\ \overset{>}{\sim}\ $}}
\def\simle{\lower0.7ex\hbox{$\ \overset{<}{\sim}\ $}}

\begin{document}

\title{Bipolaron-SO(5) Non-Fermi Liquid in a Two-channel Anderson
Model with Phonon-assisted Hybridizations}

\author{K. Hattori}
\email{hattori@issp.u-tokyo.ac.jp}
\affiliation{%
Institute for Solid State Physics, University of Tokyo, Kashiwanoha 5-1-5, Kashiwa, Chiba 277-8581, Japan
}%

\date{\today}

\begin{abstract}
We analyze non-Fermi liquid (NFL) properties  
along a line of critical points in a two-channel Anderson model with
phonon-assisted hybridizations. 
 We succeed in identifying hidden nonmagnetic SO(5) degrees of freedom
 for the valence-fluctuation regime, and we analyze the
 model on the basis of boundary conformal field theory. 
We find that the NFL spectra along the critical line, 
which is the same as those in the
 two-channel Kondo model, can be alternatively derived by a fusion in the nonmagnetic SO(5) sector.
The leading irrelevant operators near 
 the NFL fixed points vary as a function of Coulomb repulsion
 $U$; operators in the spin sector dominate for large $U$, while those
 in the SO(5) sector dominate for small $U$, and we confirm this variation in
 our numerical renormalization group calculations.
As a result, the thermodynamic singularity for small $U$ differs from
 that of the conventional two-channel Kondo problem. In particular, the impurity contribution to specific heat is
proportional to temperature and bipolaron fluctuations, which are 
coupled electron-phonon fluctuations,  
diverge logarithmically at low temperatures for small $U$.
\end{abstract}


\maketitle
\section{Introduction}\label{intro}
Strongly interacting electron-phonon systems have
attracted much attention in 
 condensed matter physics. Vibrating ion oscillations in
metal interact with conduction electrons, leading to interesting low-energy
phenomena such as superconductivity and various density-wave
states. For about three decades, Kondo effects\cite{Kondo} due to local ion oscillations
have been studied intensively by various authors.\cite{YuAnderson,Miyake,Vladar1,Vladar2,Vladar3,Fisher,Fisher2,Kusu,Yotsu,hat4level,hotta1,hotta2,hotta3} Realization of
heavy-fermion behaviors in filled-skutterudite SmOs$_4$Sb$_{12}$
under high magnetic fields\cite{SmOsSb} suggests that this compound is not a
conventional heavy-fermion system 
due to magnetic Kondo effects.\cite{Kondo} One of the distinct properties in the
filled-skutterudite structure is that the Sm atom
is located inside a large ``cage'' consisting of Sb atoms. Since the
size of the cage is much larger than that of the Sm atom, the spacial
 dependence of the potential
energy for the Sm atomic
oscillations is very shallow and the oscillations become anharmonic. Materials that have
similar cage-like structure such as clathrate compounds\cite{clathrate} and
$\beta$-pyrochlore oxides\cite{beta} have also been studied recently, partially due
to the potential application as thermoelectric materials and observation
of strong coupling superconductivity mediated by strong anharmonic oscillations.

As a pioneering work of Kondo physics in electron-phonon systems, Yu and Anderson investigated a system with one local phonon
interacting with spinless two-channel conduction electrons.\cite{YuAnderson} Vlad\'ar and
Zawadowski investigated so called two-level systems,\cite{Vladar1,Vladar2,Vladar3} where ``two-level''
represents two quasi-degenerate ion states in a double-well potential, coupled with two-channel conduction
electrons, and they proposed that it is possible to realize the two-channel Kondo
phenomena\cite{Nozieres} in this system.  There are many theories that investigate such Kondo
physics due to ionic oscillations.\cite{Fisher,Fisher2, hat4level} Recently, 
the full phase diagram of a two-channel Anderson model with
phonon-assisted hybridizations 
 was clarified\cite{Daggoto,Yashiki1,Yashiki2, UnpubHotta} 
 by using Wilson's numerical renormalization group\cite{nrg} (NRG)
method to analyze the Kondo effects in molecular systems\cite{Daggoto} and also in
cage compounds such as filled-skutterudites.\cite{Yashiki1,Yashiki2,UnpubHotta}
A line of two-channel Kondo like fixed points was found from the
 weakly correlated regime to the Kondo regime in the phase
 diagram. The effects of anharmonicity in the local
 phonon potential were also investigated.\cite{Yashiki3}

The purpose of this study is to clarify why the two
completely different regimes, {\it i.e.}, the Kondo and weak coupling
regimes 
in the model investigated in Refs. 19-21,  
can be connected smoothly along the critical
line, 
where the fixed point
spectra and the quantum numbers characterizing each of the states are 
invariant.\cite{Daggoto,Yashiki1,Yashiki2} 
In the Kondo regime,
it is natural to expect that a magnetic two-channel Kondo model (2CKM)\cite{Nozieres} describes
low-energy properties of this system. For the weak coupling regime, however, 
 it is unclear what is going on. One naive expectation is that some nonmagnetic degrees of
 freedom play important roles to realize the identical fixed point spectra
 and residual entropy\cite{Aff-ent} ln$\sqrt{2}$ with those in the
 magnetic 2CKM. 
In this respect, one can imagine 
that nonmagnetic two-channel Kondo effects occur, as had been proposed 
in two-level systems,\cite{Vladar1,Vladar2,Vladar3,Cox} 
where an ion tunnels between two local potential minima, 
interacting with two-channel conduction electrons. 
However, this expectation is not supported by the asymptotically exact
NRG results.\cite{Daggoto,Yashiki2}

In this paper, we will demonstrate that the non-Fermi liquid (NFL) of
the 2CKM can be alternatively interpreted as a nonmagnetic 
SO(5) NFL. This point of view can resolve the above
questions and we can further predict various crossover behaviors along the critical line.
 In Sec. 2, we will introduce the SO(5) degrees of freedom and develop 
a critical theory along the line of the fixed points by using
 the non-Abelian bosonization method and boundary conformal field theory. 
Section 3 will be devoted to confirming the analytic results obtained in Sec. 2 
by using Wilson's
NRG method. Finally, in Sec. 4, we will summarize
the present results.

\section{Boundary Conformal Field Theoretical Approach} 
\subsection{Two-channel Anderson Model with Phonon-assisted Hybridizations}

\subsubsection{Model}
In this paper, we investigate a two-channel Anderson
model with one-component directional local phonons that assist hybridization processes
between a localized electron at the origin and conduction
electrons.\cite{Daggoto} This model has been investigated in the context
of the Kondo effects in molecular systems while taking into account its
vibrations.\cite{Daggoto} Later, the same model was reanalyzed to
investigate the Kondo physics in cage
compounds that include magnetic ions in their cage structure.\cite{Yashiki1,Yashiki2,Yashiki3}  

The Hamiltonian is
\begin{eqnarray}
H&=&\sum_{q\sigma}\epsilon_q
(n_{qs\sigma}+n_{qp\sigma})+U(\sum_{\sigma}n_{f\sigma}-1)^2\nonumber\\
&+&\sum_{q\sigma}[f^{\dagger}_{\sigma}(V_0s_{q\sigma}+V_1xp_{q\sigma})+{\rm
 h.c.}]+\Omega b^{\dagger}b,\label{H}
\end{eqnarray}
where $s_{q\sigma}^{\dagger}(p_{q\sigma}^{\dagger})$ is the conduction
electron creation operator with the radial wavenumber $q$, the spin $\sigma$
and $s$-wave ($p$-wave) symmetry around the origin. $\epsilon_q$ represents 
the energy dispersion of the conduction electrons, and we 
set the total volume to unity. $f^{\dagger}$ is the
creation operator of a localized orbital with the spin $\sigma$ and
$n_{qs\sigma}=s_{q\sigma}^{\dagger}s_{q\sigma}$, 
$n_{qp\sigma}=p_{q\sigma}^{\dagger}p_{q\sigma}$, and 
$n_{f\sigma}=f_{\sigma}^{\dagger}f_{\sigma}$. $U$, $V_0$, and $V_1$
represent the Coulomb repulsion, the hybridization, and the
phonon-assisted hybridization, respectively. 
The model (\ref{H}) is essentially the same as the model used in
Ref. 19, in which the parameters satisfy $V_R=V_L$ and $\lambda=0$. 
The dimensionless displacement of $p$-wave local phonons is indicated by
$x=b+b^{\dagger}$ with $b^{\dagger}$ being the phonon creation
operator and $\Omega$ is the energy of the phonon. 
For simplicity, we
restrict ourselves to the particle-hole symmetric case, since the
particle-hole asymmetry does not alter our main conclusion. 

\subsubsection{Symmetry}
Before going into  detailed analysis, we list here the symmetries of
the present system. 
The Hamiltonian (\ref{H}) has three symmetries. One is the spin SU(2)
symmetry and another is the charge SU(2) symmetry. The last is
$Z_2$ symmetry, which is related to inversion symmetry at the origin. 
For each of the symmetries, there
are conserved quantities. 

For the spin SU(2) symmetry, the total spin and its
z-component are conserved. The total spin is given by
\begin{eqnarray}
{\bf S}_{\rm tot}={\bf S}+\sum_l{\bf S}_{s}(x_l) +\sum_l{\bf S}_{p}(x_l),\label{S}
\end{eqnarray}
where ${\bf S}$ is the spin of the $f$-electron and ${\bf S}_{s}(x_l)({\bf S}_{p}(x_l))$
is the spin of the $s(p)$-electron at site $x_l$ in a one-dimensional
``radial'' lattice. Since ${\bf
S}_{\rm tot}$ satisfies SU(2) commutation relations, the eigenvalue of
 ${\bf S}_{\rm tot}^2=j(j+1)$, with $j$ being half integers or integers.

With regard to the charge SU(2), it is well known that 
the total axial charge and its z-component are conserved. The axial
charge for the $f$-electron ${\bf I}=(I_x,I_y,I_z)$ is given as
\begin{eqnarray}
I_z&=&\frac{1}{2}\Big(\sum_{\sigma}n_{f\sigma}-1\Big), \label{Iz}\\
I_+&=&I_-^{\dagger}=I_x+iI_y=-f_{\uparrow}^{\dagger}f^{\dagger}_{\downarrow}.\label{Ip}
\end{eqnarray}
The axial charges for the $s$ and $p$-electrons are
defined in the same way as ${\bf S}_{s(p)}(x_l)$, and we denote them as ${\bf I}_{s}(x_l)$ and ${\bf I}_{p}(x_l)$,
respectively.\cite{axialchargeNOTE} Then, the total axial charge is given as
\begin{eqnarray} 
{\bf I}_{\rm tot}={\bf I}+\sum_l{\bf I}_{s}(x_l)+\sum_l{\bf I}_{p}(x_l).
\end{eqnarray}
The axial charge operators satisfy SU(2) commutation relations and thus, the
eigenvalue of ${\bf I}_{\rm tot}^2$ is $i(i+1)$ with $i$ being half integers or
integers.

Finally, for the $Z_2$ symmetry, the total parity $P$ is conserved:
\begin{eqnarray}
P={\rm mod}\Big(\sum_{q\sigma}n_{qp\sigma}+b^{\dagger}b,2\Big).
\end{eqnarray}
The eigenvalue of $P$ is 0 or 1. The values 0 and 1 correspond to even
 and odd parity, respectively.
Note that under the inversion operation, $s_{q\sigma}\to s_{q\sigma}$, $p_{q\sigma}\to -p_{q\sigma}$,
 and $b\to -b$. Here, since $q$ is the radial wavenumber, $q$ does not
 change.

\subsubsection{Background}
In this subsection, we explain what remains to be clarified in this
model and
what we have already understood in the previous works.\cite{Daggoto,Yashiki1,Yashiki2}  

\begin{figure}[t]
\begin{center}
    \includegraphics[width=0.45\textwidth]{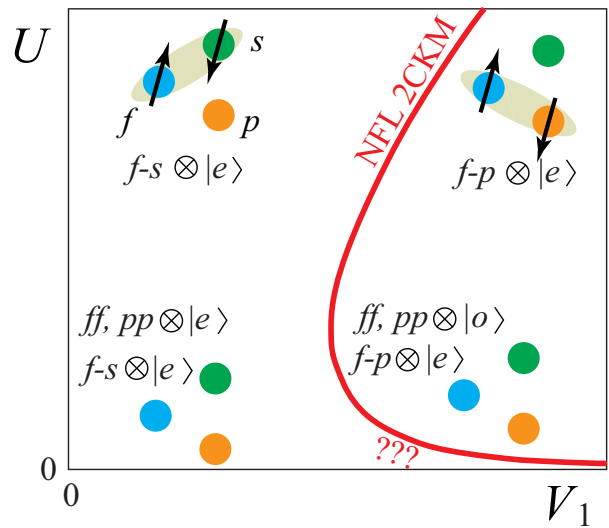}
\end{center}
\caption{(Color online) Schematic ground-state $V_1$-$U$ phase diagram of model
 (\ref{H}).\cite{Yashiki1,Yashiki2} There are two phases. One is the
 phase characterized by the parity-even ground state for small $V_1$, 
while the other is the odd-parity state for large $V_1$. Inside the each
 phase, the electron and also the phonon configurations crossover as $U$
 varies. $|e\rangle(|o\rangle)$ represents the even (odd) parity
 phonon state. Between the two phases, there is a line of NFL fixed
 points for all the values of $U>0$ where the fixed point spectra of the
 NRG\cite{Daggoto,Yashiki1,Yashiki2} are the same as
 those in magnetic 2CKM. In the figure, we depict characteristic
 configurations with the spin singlet in each region in the two
 phases. $f$-$s(p)$ indicates the spin-singlet state between $f$ and
 $s(p)$ electrons and $f\!f(pp)$ represents the double occupied state for
 the $f(p)$ electron. 
 }
\label{fig-0}
\end{figure}

As we have mentioned in Sec. 1, the global phase diagram of the
model (\ref{H}) is known in the  $V_1$-$U$ plane. We schematically draw the
phase diagram in Fig. \ref{fig-0}. There are two phases in the $V_1$-$U$
plane with fixed $\Omega$ and $V_0$.\cite{negativeU} Each phase is characterized by the
ground state parity. In the phase for small $V_1$, it is even parity,
while it is odd parity for large $V_1$. For large $U$, there appears
spin-1/2 local magnetic moment of the $f$-electron in both phases. The
spin is eventually screened via Kondo couplings by that of the $s$- or
$p$-electrons, as determined by 
the strength of $V_0$ and $V_1$. For large $U$, the phonon state can be
regarded as an 
 even-parity state denoted by $|e\rangle$ in Fig. \ref{fig-0}, which
 is continuously connected to the phonon vacuum state. Between the two
 phases, 
there is a line
of NFL fixed points characterized by the spectra equal to those in
the magnetic 2CKM.

As $U$ decreases, the local magnetic moment of the $f$-electron disappears and
the Kondo-singlet state in the spin sector gradually crossovers to the
different configurations in each of the phases. For small $V_1$ and the 
small $U$ regime, the ground state is essentially a non-interacting
state, which was called ``renormalized Fermi liquid'' in
Ref. 20. For large $V_1$ and small $U$, the ground state
consists of, in addition to the component dominant for large $U$, 
the odd-parity phonon state $|o\rangle$ coupled with 
even-parity states formed by $f$-and $p$-electrons. 
Note that near the
critical line, there are components of spin-singlet states between $f$
and $p(s)$ electrons with $|o\rangle$ for the left (right) side of the
line (not depicted in Fig. \ref{fig-0}) and their magnitudes are 
comparable with those of the spin-singlet
states with $|e\rangle$.

A natural question is that, {\it why the line of the NFL fixed point
 is continuous, even when the dominant components of the
$f$-electron in the ground states 
crossover from magnetic to nonmagnetic ones}. It should be also clarified that, {\it why even when the local magnetic moment is absent
for small $U$, the
spectra of the NFL is the same as those in the  magnetic 2CKM.} 

In Sec. \ref{BCFT}, we will construct a critical theory that 
can describe the line of the NFL fixed points. Although it is beyond the
scope of this paper 
to investigate physics very far from the critical line, it is
possible to analyze the stability of the NFL fixed points and predict
various critical behaviors in our theory.

\subsection{Non-Abelian Bosonization}\label{BCFT}
 In this section, we map the original Hamiltonian to one in an effective
 one-dimensional continuous ``radial'' space with 
only left-moving conduction electron components, and we apply 
non-Abelian bosonization.\cite{Aff1} In this approach, the free electron
part of the
Hamiltonian (\ref{H}) is written as
\begin{eqnarray}
H_0=\frac{iv_F}{2\pi}\int dx \Big[ s_{\sigma}^{\dagger}(x)\partial_x s_{\sigma}(x)+
p_{\sigma}^{\dagger}(x)\partial_x p_{\sigma}(x)
\Big],\label{H1d}
\end{eqnarray}
where $x$ is the position in the one-dimensional space and $v_F$ is the Fermi
velocity. 
\subsubsection{Conformal embedding: U(1)$\otimes$SU(2)$_2\otimes$SU(2)$_2$}

The Hamiltonian (\ref{H1d}) consists of two flavors of
conduction electrons $s$ and $p$ with the spin $j=1/2$.
When we apply non-Abelian bosonization to this model, 
the simplest way of so called conformal embedding is to bosonize, global charge U(1), spin SU(2)
and flavor SU(2) degrees of freedom.\cite{Aff2,Aff3} We set the one-dimensional
system size as $[-l,l]$ and the bosonized Hamiltonian leads 
\begin{eqnarray}
H_0&=&\frac{\pi v_F}{l}\sum_{n}\Big[\frac{1}{8}:J_nJ_{-n}:
+\frac{1}{4}:{\bf J}_n\cdot {\bf J}_{-n}: \nonumber\\
&&+\frac{1}{4}:{\bf F}_n\cdot {\bf F}_{-n}: 
\Big],\label{Hsu2}
\end{eqnarray}
where $J_n$, ${\bf J}_n=(J_n^x,J_n^y,J_n^z)$, and ${\bf
F}_n=(F_n^x,F_n^y,F_n^z)$ represent the charge, spin, and 
flavor (left moving) current operators in the Fourier space labeled by
integers $n$, respectively, and $:A:$ is the normal ordering
of operator $A$. $J_n$ satisfies the U(1) boson commutation relation, and
 ${\bf J}_n$ and ${\bf F}_n$ satisfy the SU(2)$_k$ Kac-Moody algebra with
 the level $k=2$. This conformal embedding is suitable when the
 interactions consist of, for example, exchange interactions in the spin
 sector as in the case of the conventional two-channel Kondo problem.\cite{Aff2,Aff3}
However, it is inconvenient for us, since there is no flavor SU(2)
symmetry in the Hamiltonian (\ref{H}).  

\subsubsection{Conformal embedding: SU(2)$_2\otimes$SO(5)$_2$} \label{SO5SU2}
As is well known, the symmetry in the 2CKM is higher than 
U(1)$\otimes$SU(2)$\otimes$SU(2), and it is
SU(2)$\otimes$SO(5).\cite{Aff4} 
In the NRG studies of the Hamiltonian (\ref{H}), SU(2)$\otimes$SO(5)
symmetry is also realized along the line of the NFL fixed points.\cite{Yashiki1,Yashiki2} 
 In this subsection, we clarify what SO(5) degrees of
freedom are in the model (\ref{H}).

First, we introduce the Nambu representation:
\begin{eqnarray}
\bar{\Psi}(x)&=&[s^{\dagger}_{\uparrow}(x),-p^{\dagger}_{\uparrow}(x),p_{\downarrow}(x),s_{\downarrow}(x)],\\
{\Psi}(x)&=&^t[s_{\uparrow}(x),-p_{\uparrow}(x),p^{\dagger}_{\downarrow}(x),s^{\dagger}_{\downarrow}(x)].
\end{eqnarray}
We find that the
 SO(5) ``density'' is given by using $\Psi(x)$
as 
\begin{eqnarray}
L^{ab}(x)=\sum_{\alpha=1}^4\sum_{\beta=1}^4\bar{\Psi}_{\alpha}(x)
 (\mathcal L^{ab})_{\alpha\beta} \Psi_{\beta}(x),
\end{eqnarray}
 where ${\mathcal L^{ab}}\equiv \Gamma^{ab}/2$ with $1\le a<b\le 5$ are SO(5) generators
 and defined as 4 by 4 matrices as shown in Appendix \ref{SO}.\cite{Wu} 
They define SO(5) rotations and satisfy the SO(5) commutation relation
 (\ref{f}). There are ten generators in the SO(5) group, which are 
the adjoint representation, denoted by ${\bf 10}$.
For later purposes, we define the 10-component vector ${\bf L}(x)$ as 
\begin{eqnarray}
{\bf
 L}(x)\equiv(L^{12},L^{13},L^{14},L^{15},L^{23},L^{24},L^{25},L^{34},L^{35},L^{45}), \label{tenL}
\end{eqnarray}
where $x$ dependence is omitted on the right-hand side of Eq. (\ref{tenL}). We can bosonize ${\bf L}(x)$ and 
their Fourier components ${\bf L}_n$ satisfy the following SO(5)$_k$ Kac-Moody algebra
with the level $k=2$:\cite{hat1}
\begin{eqnarray}
[{L}^{ab}_n,{L}^{cd}_m]&=& if^{ab,cd,ef}L^{ef}_{n+m}+\frac{nk}{2}\delta_{ac}\delta_{bd}\delta_{n+m,0}.
\end{eqnarray}
Here, $f^{ab,cd,ef}$ is the SO(5) structure constant and is given by
Eqs. (\ref{f0}) and (\ref{f}). 

The Hamiltonian (\ref{H1d}) is bosonized by
using the conformal embedding SU(2)$_2\otimes$SO(5)$_2$, leading to 
\begin{eqnarray}
H_0=\frac{\pi v_F}{l}\sum_{n}\Big[\frac{1}{4}:{\bf J}_n\cdot {\bf J}_{-n}:
 +\frac{1}{8}:{\bf L}_n\cdot {\bf L}_{-n}: 
\Big]. \label{H2}
\end{eqnarray}
Indeed, the central charge for the SO(5) sector is $c_{SO(5)}=5/2$ and that
for the spin sector is $c_{s}=3/2$, leading to $c_s+c_{SO(5)}=4$ as it
should be. This form (\ref{H2}) is used in a spin-3/2 dipole-octupole
Kondo problem.\cite{hat1} There, $\Psi(x)$ corresponds to the spin-3/2 fermion
operator and the SO(5) generators correspond to linear
combinations of the spin-3/2 dipole and octupole operators, and the spin
current ${\bf J}$ is replaced by the SU(2) axial charge current. Since, in the spin-3/2 model, quadrupole
operators are classified as the SO(5) vector, i.e., the ${\bf 5}$ representation, we
define the corresponding degrees of freedom in our model and they are
given by
\begin{eqnarray}
n^{a}(x)=\frac{1}{2}\sum_{\alpha=1}^4\sum_{\beta=1}^4\bar{\Psi}_{\alpha}(x)
 (\Gamma^{a})_{\alpha\beta} \Psi_{\beta}(x),
\end{eqnarray}
where $\Gamma^a$'s with $1\le a\le5$ are 4 by 4 matrices defined in
Appendix \ref{Gammamatrix}. In the following, we use a five-component
vector ${\bf n}(x)=(n^1(x),\cdots,n^5(x))$.
Note that the spin eigenvalue of ${\bf n}(x)$ is $j=1$ and 
the z-component $j_z=0$, while $j=0$ for ${\bf L}(x)$. $j_z=\pm 1$
components of the ${\bf 5}$ representation ${\bf n}_{\pm}(x)$ 
can be constructed 
by applying the spin operators ${J}^{\pm}(x)={J}^{x}(x)\pm i{J}^{y}(x)$ to ${\bf n}(x)$.
In total, 28  operators, ${\bf J}$, ${\bf L}$, ${\bf n}$, and ${\bf n}_{\pm}$, form a
complete set of the conduction electron ``density'' operators
 in the sense of the Nambu representation.

It is also important to check the ``independence'' of the two sectors and we
find, by direct calculations, $[L^{ab}(x), J^{\mu}(x')]=0$, {\it i.e.}, they are ``independent.''
The actual form of ${\bf L}(x)$ and ${\bf n}(x)$ is given in Table
\ref{tbl-0}. It is clear that ${\bf L}(x)$ consists of charge, flavor, and
spin-singlet pairing operators, while ${\bf n}(x)$ consists of spin-triplet
pairing and a complex object of spin-flavor operators, see Appendix \ref{AxialC}. 
In terms of $L^{ab}_n$, the charge current $J_n$ and the flavor current
${\bf F}_n$ in
Eq. (\ref{Hsu2}) are
related to $-L^{15}_n$, and $(-L^{34}_n,L^{24}_n,L_n^{23})$, respectively.

By using this conformal embedding, the free electron
energy spectra $E_0$ are calculated via eigenvalues of the Casimir operators in
the two sectors and are given by\cite{hat1}
\begin{eqnarray}
E_0=\frac{\pi v_F}{l}\Big[\frac{j(j+1)}{4}+\frac{C_{SO(5)}}{8}+m\Big]. \label{eigenE}
\end{eqnarray}
Here, $m$ is a non-negative integer, $m\ge 0$ and $C_{SO(5)}$ is the
eigenvalue of the Casimir operator in the SO(5) sector. The eigenvalue
depends on the irreducible representations and is given as $C_{SO(5)}=0$
for the identity ${\bf 1}$, $5/2$ for the spinor ${\bf 4}$, and $4$ for
the vector ${\bf 5}$. Eigenstates with $m=0$ means that the eigenstate
is a primary state,
while those with $m\ge 1$ indicates that the states include particle-hole excitations
and are classified as descendant states in the conformal tower characterized
by a set of primary states.\cite{cft} 
The SU(2)$_2$ Kac-Moody algebra restricts possible values of $j$ for the
primary states in the spin sector. 
The spin $j$ of the primary states should be $0\le j\le k/2=1$.\cite{Aff2}
The SO(5)$_2$ Kac-Moody algebra also restricts
the number of primary states in the SO(5) sector, and, as shown in Appendix
\ref{primary}, there are 
three primary states in the SO(5) sector: the identity ${\bf 1}$, the spinor ${\bf
4}$, and the vector ${\bf 5}$. 
Using Eq. (\ref{eigenE}), we can reproduce
the free-electron spectra as shown in Table \ref{tbl-1} (a).

\begin{table}[!t]
\caption{List of operators in the SO(5) sector. Each of irreducible
 representation is labeled by $(j_{j_z}$, dimension of the SO(5) group). 
Position index $x$ is omitted for ${\bf L}(x)$ and ${\bf n}(x)$ and the
 operator forms shown are those multiplied by a factor 2: 
 $2{\bf L}(x)$, $2{\bf n}(x)$, $2{\bf L}_I$ and $2{\bf n}_I$.
  Each operator is labeled by the parity
 $P$, the axial charge eigenvalue $i$ and the z-component
 $i_z$, and the flavor $F$ (see, analysis in Sec. \ref{mini}). 
$i_{i_z}=1_{\pm 1}$ means that the operator is in linear combinations of
 $i_z=1$ and $-1$.}

   \begin{tabular}{cccccc}
\hline
\hline
$(j_{j_z},{\rm SO(5)})$ & label & operator form & $P$ & $i_{i_z}$ & $F$\\
\hline
 (0$_0$,{\bf 10}) &      $L^{12}$            &
	    $-s^{\dagger}_{\uparrow}s^{\dagger}_{\downarrow}+p^{\dagger}_{\uparrow}p^{\dagger}_{\downarrow}+{\rm
	    h.c.}$ & 0 & $1_{\pm 1}$  & 1 \\
  &      $L^{13}$            & $i(s^{\dagger}_{\uparrow}s^{\dagger}_{\downarrow}+p^{\dagger}_{\uparrow}p^{\dagger}_{\downarrow})+{\rm
	    h.c.}$ & 0 &$1_{\pm 1}$  & 0\\
  &      $L^{14}$            & $-s^{\dagger}_{\uparrow}p^{\dagger}_{\downarrow}-p^{\dagger}_{\uparrow}s^{\dagger}_{\downarrow}+{\rm
	    h.c.}$ & 1 &$1_{\pm 1}$  &0\\
  &      $L^{15}$            &
	    $-\sum_{\sigma}(s^{\dagger}_{\sigma}s_{\sigma}+p^{\dagger}_{\sigma}p_{\sigma})+2$
	    & 0 &$1_{0}$  & 0\\
  &      $L^{23}$            &
	    $\sum_{\sigma}(s^{\dagger}_{\sigma}s_{\sigma}-p^{\dagger}_{\sigma}p_{\sigma})$
	    & 0 &$1_{0}$  & 1\\
  &      $L^{24}$            &
	    $-i\sum_{\sigma}(s_{\sigma}^{\dagger}p_{\sigma}-p^{\dagger}_{\sigma}s_{\sigma})$
	    & 1 &$0_0$ & 1 \\
  &      $L^{25}$            & $i(s^{\dagger}_{\uparrow}s^{\dagger}_{\downarrow}-p^{\dagger}_{\uparrow}p^{\dagger}_{\downarrow})+{\rm
	    h.c.}$ & 0 &$1_{\pm 1}$  & 1\\
  &      $L^{34}$            &
	    $-\sum_{\sigma}(s_{\sigma}^{\dagger}p_{\sigma}+p^{\dagger}_{\sigma}s_{\sigma})$
	    & 1 &$1_{0}$  &0 \\
  &      $L^{35}$            & $s^{\dagger}_{\uparrow}s^{\dagger}_{\downarrow}+p^{\dagger}_{\uparrow}p^{\dagger}_{\downarrow}+{\rm
	    h.c.}$ & 0 &$1_{\pm 1}$  & 0\\
  &      $L^{45}$            & $i(s^{\dagger}_{\uparrow}p^{\dagger}_{\downarrow}+p^{\dagger}_{\uparrow}s^{\dagger}_{\downarrow})+{\rm
	    h.c.}$ & 1 &$1_{\pm 1}$  & 0\\
\hline
 (1$_0$,{\bf 5}) &      $n^{1}$            &
    $i(s^{\dagger}_{\uparrow}p^{\dagger}_{\downarrow}+s^{\dagger}_{\downarrow}p^{\dagger}_{\uparrow})
    +  {\ \rm h.c.}$ & 1 &$1_{\pm 1}$  & 1\\
  &      $n^{2}$            &
	    $\sum_{\sigma}\sigma(s^{\dagger}_{\sigma}p_{\sigma}+p^{\dagger}_{\sigma}s_{\sigma})$
	    & 1 &$0_{0}$  & 0\\
  &      $n^{3}$            &
	    $-i\sum_{\sigma}\sigma(s^{\dagger}_{\sigma}p_{\sigma}-p^{\dagger}_{\sigma}s_{\sigma})$
	    & 1 &$1_0$  & 1\\
  &      $n^{4}$            & $\sum_{\sigma}\sigma(s^{\dagger}_{\sigma}s_{\sigma}-p^{\dagger}_{\sigma}p_{\sigma})$ & 0 &$0_{0}$ &1 \\
  &      $n^{5}$            &
	    $-s^{\dagger}_{\uparrow}p^{\dagger}_{\downarrow}-s^{\dagger}_{\downarrow}p^{\dagger}_{\uparrow}+{\
	    \rm h.c.}$ & 1 &$1_{\pm 1}$ &1 \\
\hline
 (0$_0$,{\bf 10}) &      $L^{12}_I$            &
	    $\tau_z(f^{\dagger}_{\uparrow}f^{\dagger}_{\downarrow}+f_{\downarrow}f_{\uparrow})$
	    & 0 &$1_{\pm 1}$  & 1\\
  &      $L_I^{13}$            &
	    $-i(f^{\dagger}_{\uparrow}f^{\dagger}_{\downarrow}-f_{\downarrow}f_{\uparrow})$
	    & 0 &$1_{\pm 1}$  &0\\
  &      $L_I^{14}$            &
	    $-\tau_x(f^{\dagger}_{\uparrow}f^{\dagger}_{\downarrow}+f_{\downarrow}f_{\uparrow})$
	    & 1 &$1_{\pm 1}$ &0 \\
  &      $L_I^{15}$            & $-(\sum_{\sigma}n_{f\sigma}-1)$ & 0 &$1_{0}$ & 0\\
  &      $L_I^{23}$            & $\tau_z(\sum_{\sigma}n_{f\sigma}-1)$ & 0 &$1_{0}$ &1 \\
  &      $L_I^{24}$            & $-\tau_y{\mathcal P}$ & 1 &$0_{0}$  & 1\\
  &      $L_I^{25}$            &
	    $-i\tau_z(f^{\dagger}_{\uparrow}f^{\dagger}_{\downarrow}-f_{\downarrow}f_{\uparrow})$
	    & 0 &$1_{\pm 1}$  & 1\\
  &      $L_I^{34}$            & $\tau_x(\sum_{\sigma}n_{f\sigma}-1)$ & 1 &$1_{0}$ &0 \\
  &      $L_I^{35}$            &
	    $-(f^{\dagger}_{\uparrow}f^{\dagger}_{\downarrow}+f_{\downarrow}f_{\uparrow})$
	    & 0 &$1_{\pm 1}$  & 0\\
  &      $L_I^{45}$            & $i\tau_x(f^{\dagger}_{\uparrow}f^{\dagger}_{\downarrow}-f_{\downarrow}f_{\uparrow})$ & 1 &$1_{\pm 1}$ &0 \\
\hline
 (0$_0$,{\bf 5}) &  $n_I^{1}$            &
	    $-\tau_y(f^{\dagger}_{\uparrow}f^{\dagger}_{\downarrow}+f_{\downarrow}f_{\uparrow})$
	    & 1 &$1_{\pm 1}$  &1\\
  &      $n_I^{2}$            &
	    $\tau_x{\mathcal P}$ &1 &$0_{0}$ &0 \\
  &      $n_I^{3}$            & $\tau_y(\sum_{\sigma}n_{f\sigma}-1)$ & 1 &$1_{0}$  &1\\
  &      $n_I^{4}$            & $\tau_z{\mathcal P}$ & 0 &$0_{0}$  &1\\
  &      $n_I^{5}$            &
	    $-i\tau_y(f^{\dagger}_{\uparrow}f^{\dagger}_{\downarrow}-f_{\uparrow}f_{\downarrow})$
	    & 1 &$1_{\pm 1}$ &1 \\
\hline
\hline
   \end{tabular}
\label{tbl-0}
\end{table}

\subsubsection{Local operators classified in a hidden SO(5) group}\label{SO5local}
In this subsection, 
we will introduce local ``flavor'' degrees of freedom.
Using them combined with ${\bf I}$, we will show that we can construct local SO(5) degrees of
freedom in terms of local operators: $f$, $f^{\dagger}$, $b$, and
$b^{\dagger}$ for small $U$.

In Ref. 21, it is shown that the energy spectra of
small-cluster 
problems capture the essential aspect of the critical line
obtained in the NRG calculations. Namely, there is a level crossing of the ground states
of the $(j,i)=(0,1/2)$ sector. 
Here, $j(i)$ represents the eigenvalue of
total spin (axial charge). 
Let us briefly explain their results in the following. 

First, for large $V_1$, it is natural to consider a two-site problem,
where 
$f$ and $p(0)$ electrons and $b$ are taken into account. For $U=0$, the Lang-Firsov
transformation\cite{Lang} provides the exact solution of this problem. The ground
state for $U=0$ is doubly degenerate. The two are  
two-electron states with the spin being singlet. This degeneracy is
distinguished by the parity, {\it i.e.}, one is an even-parity state $|\phi_e\rangle$ and the
other is odd: $|\phi_o\rangle$.
This degeneracy, however, is lifted by $U$, and for small but finite $U$,
the ground state is the even-parity state with a small excitation gap to the
odd-parity state. Importantly, in the low-energy states 
for $U=0$, the phonon $b$ appears only in the special 
combinations of the even- and odd-parity states:\cite{Yashiki2}
\begin{eqnarray}
|e\rangle&=&\cosh\Big[2\lambda(b-b^{\dagger})\Big]|0\rangle_{\rm ph},\\
|o\rangle&=&\sinh\Big[2\lambda(b-b^{\dagger})\Big]|0\rangle_{\rm ph}.
\end{eqnarray}
Here, $|0\rangle_{\rm ph}$ represents the vacuum of the phonon and $\lambda=-V_1/\Omega$.
The other phonon states are in higher energy above $\Omega$, and thus do
not play important roles.

The three-site problem, where $f$, $p(0)$, and $s(0)$ electrons and the
phonons $b$ are taken into account, exhibits a qualitatively correct phase diagram, if we see the
sector of two-electron or four-electron states, {\it i.e.}, $i=1/2$ states.
For small $V_1$, the ground state of the two-electron sector 
is an even-parity spin-singlet state, 
$|\phi'_e\rangle\sim |\phi_e\rangle$,
 while for large $V_1$
the ground state changes to an odd-parity spin-singlet state  $|\phi'_o\rangle\sim
|\phi_o\rangle$. The same is true in the four-electron sector, and we denote
them as $|\psi'_e\rangle$ and $|\psi'_o\rangle$. Their spin eigenvalues
are also
$j=0$. Thus, there is a line where a level crossing occurs.
Along the
level-crossing line, the energies of these four states---$|\phi'_e\rangle$, $|\phi'_o\rangle$, $|\psi'_e\rangle$, and
$|\psi'_o\rangle$---coincide and they form an SO(5) spinor, $\bf 4$
representation, in the SO(5) group. With regard to the $(i,j)=(0,1/2)$ sector, there are
quasi-degenerate states with even and odd parity that originate in
the ground states for the two-site problem with one electrons in the $s$-orbital.
One is an SO(5)-singlet, while the other is one of the states
in the ${\bf 5}$ representation.

Our assumption on the phonon degrees of freedom for small $U$ 
is that the only two states, such as 
$|e\rangle$ and $|o\rangle$, which are solely constructed by the phonon
part,  are important in the low-energy
properties along the critical line. Indeed, this is valid in the limit of
$U=0$, since the first excited state lies at $\Omega$, and thus it does not
play any role in the low-energy physics.\cite{Yashiki2} 
Then, we can
construct ``flavor'' operators characterized by the Pauli matrices in
these two bases and we denote them as $\tau_x,\tau_y$, and $\tau_z$,
identifying the even-parity state as $|\!\!\Uparrow\rangle$ and the odd-parity state
as $|\!\Downarrow\rangle$. 
The level crossing in the three-site problem is described by the
presence of a term $\propto (U-U_c)\tau^z$ while fixing other parameters, where $U_c$ is
the level-crossing value of $U$. Note that, for $U>0$, there is always a finite gap between the even
and odd ground states in the (0,1/2) sector,  since the
gap arises mainly from the fact that  the two are classified in a different
irreducible representation and thus have generally different energies.

Another local degree of freedom is the $f$ operator part. We 
consider quadratic operators in terms of $f_{\sigma}$ and $f^{\dagger}_{\sigma}$. 
Possible linear combinations are the spin ${\bf S}$ and the axial charge ${\bf
I}$; see Eqs. (\ref{S}), (\ref{Iz}) and (\ref{Ip}). Since the SO(5)
sector is nonmagnetic, 
 we consider the $f$-electron part of the axial charge, ${\bf
I}$, in detail.

 ${\bf I}$ satisfies the SU(2) commutation relations, and thus, 
the eigenvalue of
${\bf I}^2$ is $i(i+1)$ with $i=0$ or $1/2$. 
When $I_{\alpha}$ acts on an $i=0$ subspace 
$f^{\dagger}_{\sigma}|0\rangle$, where $|0\rangle$ is the vacuum of the
$f$-electrons, $I_{\alpha}f^{\dagger}_{\sigma}|0\rangle=0$. 
Because of  this
property, the operation of $I_{\alpha}$ is automatically projected on an 
$i=1/2$ subspace, $|0\rangle$ and $f_{\uparrow}^{\dagger}f_{\downarrow}^{\dagger}|0\rangle$.
This is very important in the derivation of local SO(5) degrees of
freedom below, 
since the algebra of $\tilde{I}_{\alpha}\equiv 2I_{\alpha}$ is very similar to that of the Pauli
matrices. Indeed, $\tilde{I}_{\alpha}$'s satisfy
\begin{eqnarray}
[\tilde{I}_{\alpha},\tilde{I}_{\beta}]&=&2i\epsilon_{\alpha\beta\gamma}\tilde{I}_{\gamma},\\
\{\tilde{I}_{\alpha},\tilde{I}_{\beta}\}&=&2{\mathcal P}\delta_{\alpha\beta},\\
{\mathcal P}&\equiv&\tilde{I}_x^2=\tilde{I}_y^2=\tilde{I}_z^2=2n_{f\uparrow}n_{f\downarrow}-\tilde{I}_z,\\
\tilde{I}_{\alpha}{\mathcal P}&=&{\mathcal P}\tilde{I}_{\alpha}=\tilde{I}_{\alpha},
\end{eqnarray}
where $\mathcal P$ is the projection operator onto the $i=1/2$
subspace and is alternatively given by ${\mathcal P}=4{\bf I}^2/3$.

Now, let us first introduce an SO(5) vector, {\it i.e.},  the ${\bf 5}$
representation ${\bf n}_I$:
\begin{eqnarray}
{\bf n}_I=\frac{1}{2}(
\tilde{I}_x\tau_y, {\mathcal P}\tau_x,\tilde{I}_z\tau_y,{\mathcal P}\tau_z,-\tilde{I}_y\tau_y
). \label{nI}
\end{eqnarray}
Note that ${\bf n}_I$ is a spin-singlet operator unlike ${\bf n}(x)$ of the
conduction electron with the spin being triplet, and ${\bf n}_I$ includes $\tilde{I}_{\alpha}$ in
all the components, which  means ${\bf n}_I$ acts on the $i=1/2$
subspace. At this stage, it is not proved that ${\bf n}_I$ transforms as
an SO(5) vector, since we have not
defined any local SO(5) generator. Thus, let us define a trial local SO(5)
generators, ${\bf 10}$, via 
Eq. (\ref{nGamma}), 
$2L_I^{ab}\equiv[2n^a_I,2n^b_I]/(2i)$, and we obtain
\begin{eqnarray}
{\bf L}_I&=&\frac{1}{2}(
-\tilde{I}_x\tau_z, -\tilde{I}_y, \tilde{I}_x\tau_x, -\tilde{I}_z,
\tilde{I}_z\tau_z, -{\mathcal P}\tau_y, -\tilde{I}_y\tau_z,\nonumber\\
&&\ \ \tilde{I}_z\tau_x, \tilde{I}_x, \tilde{I}_y\tau_x).\label{LI}
\end{eqnarray}
The trial SO(5) generators $L_I^{ab}$'s, indeed,   
satisfy the SO(5) commutation relations (\ref{f}):
$[L_I^{ab}, L_I^{cd}]=-i(\delta_{bc}
L_I^{ad}-\delta_{ac} L_I^{bd}-\delta_{bd}
L_I^{ac}+\delta_{ad} L_I^{bc})$. Thus, $L^{ab}_I$ denotes the SO(5)
generators that we seek. 
We can also confirm that the commutation relation between ${2\bf
L}_I$ and ${2\bf n}_I$ is similar to Eq. (\ref{CommLG}), 
$
[2n_I^a,2L_I^{bc}]=-i({\rm sgn}(c-a)\delta_{ab}n_I^c
+{\rm sgn}(b-a)\delta_{ac}n_I^b)
$
{\it i.e.}, 
${\bf n}_I$ is the ${\bf 5}$ representation of the local 
SO(5) group defined by ${\bf L}_I$.

Finally, we note that  the four states 
\begin{eqnarray}
^t(|\!\!\Downarrow\rangle,
 -|\!\!\Uparrow\rangle,-f^{\dagger}_{\uparrow}f^{\dagger}_{\downarrow}|\!\!\Uparrow\rangle,f^{\dagger}_{\uparrow}f^{\dagger}_{\downarrow}|\!\!\Downarrow\rangle)
\end{eqnarray}
 transform as an SO(5) spinor, ${\bf 4}$ representation. This can be
 checked by applying ladder operators in the SO(5) group listed in
 Appendix \ref{primary}.

\subsubsection{Exchange interactions}\label{mini}

So far, we have analyzed how SO(5) degrees of freedom appear in the free
part of the Hamiltonian (\ref{H1d}) and what local SO(5) operations for
small $U$ are. In this subsection, we phenomenologically discuss possible and/or impossible
interactions under the SU(2)$\otimes$SO(5) symmetry, which are closely
related to two fusions introduced in Sec. \ref{fus}. 
We show two possible exchange interactions but they are 
not the ``microscopic'' effective Hamiltonian of
(\ref{H}). They should be regarded as one of the effective interactions for
the coarse-grained system.

 The simplest invariant form under SU(2)$\otimes$SO(5) symmetry is exchange
 interactions between the impurity and the conduction electron component at the origin $x=0$. 
Spin-spin exchange interactions are evidently possible under
SU(2)$\otimes$ SO(5) symmetry:
\begin{eqnarray}
H_{\rm s} = J{\bf S}\cdot{\bf J}(0),  \label{Heff2_S}
\end{eqnarray}
with $J$ being the coupling constant. This term preserves the
symmetry of the original Hamiltonian (\ref{H}), and can be regarded
as the effective interactions after integration of the phonon and ${\bf
I}^2=3/4$ sector. 

Similarly, exchange interactions in the SO(5) sector are also possible:
\begin{eqnarray}
H_{\rm SO(5)} = K  {\bf L}_I\cdot{\bf L}(0). \label{Heff2}
\end{eqnarray}
Here,  $K$ denotes the coupling constants. This can be obtained after integrating the sector of ${\bf I}^2=0$.
Note that due to the mismatch in the eigenvalue of the
spin $j$, a term ${\bf n}_I\cdot {\bf n}(0)$ cannot exist, when the spin
SU(2) or time reversal symmetry is present. 


In order to understand that the interaction (\ref{Heff2}) respects 
the original symmetry of the Hamiltonian (\ref{H}), namely the charge SU(2) symmetry,
 the spin SU(2) symmetry, and the $Z_2$ symmetry, 
 we expand ${\bf L}_I\cdot {\bf L}(0)$ in the fermion representations:
\begin{eqnarray}
{\bf L}_I\cdot{\bf L}(0)&=&
\frac{1}{2}\Big\{\Big[I_+ \Big(
s_{\downarrow}(0)s_{\uparrow}(0)+p_{\downarrow}(0)p_{\uparrow}(0)
\Big)+{\rm h.c.}\Big]\nonumber\\
&+&
I_z\Big[
\sum_{\sigma}\Big(s^{\dagger}_{\sigma}(0)s_{\sigma}(0)
+p^{\dagger}_{\sigma}(0)p_{\sigma}(0)\Big)-2
\Big]\Big\}\nonumber\\
&+&
\frac{1}{2}\tau_z\Big\{\Big[I_+\Big(
s_{\downarrow}(0)s_{\uparrow}(0)-p_{\downarrow}(0)p_{\uparrow}(0)
\Big)+{\rm h.c.}\Big]\nonumber\\
&+&
I_z\sum_{\sigma}\Big[s^{\dagger}_{\sigma}(0)s_{\sigma}(0)
-p^{\dagger}_{\sigma}(0)p_{\sigma}(0)\Big]\Big\}\nonumber\\
%
%
&-&
\frac{1}{2}\tau_x\Big\{\Big[
I_+\Big(
p_{\downarrow}(0)s_{\uparrow}(0)+s_{\downarrow}(0)p_{\uparrow}(0)
\Big)+{\rm h.c.}\Big]\nonumber\\
&+&
I_z\sum_{\sigma}\Big[
p^{\dagger}_{\sigma}(0)s_{\sigma}(0)+{\rm h.c.}
\Big]\Big\}\nonumber\\
&+&
\frac{i}{4}\tau_y{\mathcal P}\sum_{\sigma}\Big[
s^{\dagger}_{\sigma}(0)p_{\sigma}(0)-{\rm h.c.}
\Big],\\
&\equiv&
{\bf I}\cdot \Big[{\bf I}(0)
+\tau_z {\bf I}'(0)
-\tau_x {\bf I}''(0)\Big]-\frac{1}{2}\tau_y{\mathcal P\mathcal D}(0).\nonumber\\
\label{II}
\end{eqnarray}
Here, ${\bf I}(0)$ is the local axial charge for the conduction
electrons. As shown in Appendix \ref{AxialC}, ${\bf I}'(0)$ is the local longitudinal-flavor axial charge,
 while ${\bf I}''(0)$ is the local transverse-flavor axial charge. 
All three transform as vectors under the charge SU(2) rotations,
while ${\mathcal D}(0)$ is  
invariant; see Appendix \ref{AxialC}. Since ${\bf I}$ transforms as a vector and $\mathcal P$ is invariant
under the charge SU(2) rotations, Eq. (\ref{II}) is invariant under the
charge SU(2) operations. As for the parity, ${\bf I}$,  
$\tau_z$, $\mathcal P$, ${\bf I}(0)$, and ${\bf I}'(0)$ are even parity, while 
$\tau_x$, $\tau_y$, ${\bf I}''(0)$ and ${\mathcal D}(0)$  are odd
parity. 
Thus, Eq. (\ref{II}) is even parity, {\it i.e.}, invariant under the
inversion operation. Finally, since all the terms are spin singlet,
Eq. (\ref{II}) is invariant under the spin SU(2) operations. These facts
confirm that 
${\bf L}(0)\cdot {\bf L}_I$ is invariant under the original symmetry.


In addition to the symmetry of the Hamiltonian (\ref{H}), the exchange interactions
 (\ref{Heff2_S}) and (\ref{Heff2}) have an additional flavor symmetry. The flavor
symmetry operation is defined as $|\!\!\Uparrow\rangle \!\leftrightarrow\!
|\!\!\Downarrow\rangle$ and $s_{\sigma}\! \leftrightarrow\!
p_{\sigma}$. There are two kinds of operators in terms of the flavor symmetry:
even or odd. Even-flavor operators are denoted by $F=0$, while
odd ones are denoted by $F=1$ in Table \ref{tbl-0}. The values of $F$ are related to
the eigenvalues of the flavor transformation as $(-1)^F$.
As will be investigated in Sec. \ref{stability}, this symmetry  breaking 
drives the system away from the critical line.

Note that there is also flavor SU(2) symmetry
in Eq. (\ref{Heff2}), when we define the local flavor operators 
as ${\bf F}_I\equiv$$(-L_I^{34},L_I^{24},L_I^{23})$. 
It is also important to note that the phonon flavor operator
$\vec{\tau}$ 
cannot directly couple with conduction electron flavor ${\bf F}_n$ in the presence of
the charge SU(2) symmetry. This means that simple flavor-exchange Kondo
couplings 
\begin{eqnarray}
  H_{\rm flavor}=J_f\vec{\tau}\cdot {\bf F}(0)
\end{eqnarray}
 never appear under the presence of 
particle-hole symmetry and $H_{\rm flavor}$ is
 absent also away from the critical line. In this sense, even away from the
critical line, the screening
processes of impurity degrees of freedom in the model (\ref{H}) with the
charge SU(2) symmetry are not the same
as those in the flavor Kondo model such as in the two-level
systems\cite{Cox}. Away from the critical line with the charge SU(2)
symmetry, 
a possible form in the (phonon-only) flavor
 interaction is Ising like:
\begin{eqnarray}
  H_{\rm flavor}^{xy}=(J^x_f\tau_x + J^y_f\tau_y )F^y(0) 
   =(J_f^x\tau_x + J_f^y\tau_y ){\mathcal D}(0).
\end{eqnarray}
This is because only $\mathcal D(0)=F^y(0)=L^{24}(0)$ is a charge and
also a 
spin SU(2) singlet operator in the ``local density'' form constructed by
the conduction electron operators. Since this is an Ising interaction,
when only this term is present, nonmagnetic ``Kondo effects'' never occur.


\begin{table}[!b]
\caption{(a) Spectra of the free Hamiltonian (\ref{H1d}) for a
 non-degenerate ground state. (b) The spectra
 at the NFL fixed point. The energy $E_0$ and $E$ are measured in the unit of $\pi
 v_F/l$. (c) Operator contents at the NFL fixed point. $\Delta$ is the
 scaling dimension of the operators labeled by the quantum numbers $j$
 and the dimension of the irreducible representation in the SO(5) group.}
   \begin{tabular}{ccc|ccc|ccc}
\hline
\hline
\multicolumn{3}{c|}{(a)} &\multicolumn{3}{c|}{(b)}
    &\multicolumn{3}{c}{(c)}\\
         $\ \ j\ \ \ $& SO(5) & $\ \ \ E_0\ \ \ $ &$\ \ j\ \ \ $ &
    SO(5) & $\ \ \ E\ \ \ $ &$\ \ j\ \ \ $&SO(5)&$\ \ \ \Delta\ \ \ $\\
\hline

       0            & ${\bf 1}$  & $0$            &$\frac{1}{2}$ & ${\bf
		    1}$ & 0 & 0& ${\bf 1}$ & 0\\
       $\frac{1}{2}$& ${\bf 4}$  & $\frac{1}{2}$  &0 & ${\bf 4}$
		    &$\frac{1}{8}$ & 0 & ${\bf 5}$ & $\frac{1}{2}$          \\
       1            & ${\bf 5}$  & $1$            &$\frac{1}{2}$ & ${\bf
		    5}$ &$\frac{1}{2}$ & 1 & ${\bf 1}$& $\frac{1}{2}$\\
       0            & ${\bf 10}$ & $1$            &1 & ${\bf 4}$ &
			$\frac{5}{8}$ &$\frac{1}{2}$ & ${\bf 4}$&$\frac{1}{2}$\\
       1            & ${\bf 1}$  & $1$            &$\frac{3}{2}$ &${\bf
		    1}$ &$1$ &$\frac{1}{2}$&${\bf 4}$& $\frac{1}{2}$\\
       $\frac{3}{2},\frac{1}{2}$            & ${\bf 4}$&   $\frac{3}{2}$
	    &$\frac{1}{2}$ &${\bf 10}$ &$1$ & 1& ${\bf 5}$ & $1$\\
   $\frac{1}{2}$                & ${\bf 16}$&      $\frac{3}{2}$
	    &$\frac{1}{2}$ &${\bf 1}$ &$1$ & 0& ${\bf 10}$ & 1\\
\hline
\hline
   \end{tabular}
\label{tbl-1}
\end{table}

\subsubsection{Fusions} \label{fus}
In this subsection, we introduce two different fusions\cite{Aff2,Aff3} 
that derive the spectra
obtained in the NRG calculations along the critical
line.\cite{Daggoto,Yashiki1,Yashiki2} Although, due to the fact that the model
(\ref{H}) is not described by a simple exchange Hamiltonian, we cannot 
carry out a direct impurity absorption as in exchange models such as multi-channel and
spin-3/2 multipolar Kondo problems\cite{Aff2,Aff3,hat1}, 
we will show that two fusions indeed derive the same NFL spectra as in the 2CKM. 

The first candidate of the fusion is
spin-$1/2$ fusion in the SU(2)$_2$ sector, which is the same as the case
of 2CKM\cite{Aff2,Aff3}, leading to the NFL spectra
shown in Table \ref{tbl-1} (b). This fusion is physically natural and
easy to understand, when we consider the process from large $U$ limit, since for
large $U$ the relevant operator is expected to be the spin ${\bf S}$. 

We find that there is an alternative way to derive the same NFL
spectra, i.e., SO(5) spinor ${\bf 4}$ fusion:\cite{hat1}
$^t(|\!\!\Downarrow\rangle,
 -|\!\!\Uparrow\rangle,-f^{\dagger}_{\uparrow}f^{\dagger}_{\downarrow}|\!\!\Uparrow\rangle,f^{\dagger}_{\uparrow}f^{\dagger}_{\downarrow}|\!\!\Downarrow\rangle)
$. Since primary states in the
 SO(5)$_2$ sector are ${\bf 1}$, ${\bf 4}$, or ${\bf 5}$, the fusion rule
 is ${\bf 1}\to {\bf 4}$, ${\bf 4}\to {\bf 1}\oplus {\bf
 5}$, and ${\bf 5}\to {\bf 4}$. These are obtained by discarding ${\bf
 10}$ and ${\bf 16}$ representations in the SO(5) direct products: ${\bf
 4}\otimes{\bf 4}={\bf 1}\oplus {\bf 5}\oplus {\bf 10}$ and ${\bf
 5}\otimes {\bf 4}={\bf 4}\oplus {\bf 16}$. 
Indeed, this fusion rule generates the same low-energy spectra as
the spin-1/2 fusion, as shown in Table \ref{tbl-1}(b). 

As for double fusions\cite{Aff3} of
spin $1/2$ and SO(5) ${\bf 4}$, the two different fusions lead to  
 the identical operator content\cite{Lud,hat1} shown in Table \ref{tbl-1} (c).
All the operators in Table \ref{tbl-1} (c) that satisfy the
SU(2)$\otimes$SO(5) symmetry 
can be present along the critical line. 
We expect that, 
for large $U$, the dominant leading irrelevant operator is in the spin
sector, while for small $U$, it is in the SO(5) sector, since ${\bf S}$
is not active for small $U$. 

Now, one might wonder what the difference between the two fusions is. 
We consider that these two are simply equivalent. 
In order to understand this,  we show an example
of this kind of situation in an impurity Anderson model, when the Coulomb
interaction $U$ varies from $\infty$ to $0$ while maintaining particle-hole symmetry. 

As is well known, the ground-state spectra of the Anderson model from 
 the strong to the weak coupling regime are the same as 
 those in the Kondo model with the charge quantum number being shifted. The spectra in the Kondo model 
are obtained by a 
spin-1/2 fusion via direct spin absorption.\cite{Aff1} Alternatively, the same spectra can be
obtained by an axial charge $i=1/2$ fusion, {\it i.e.,} a $\pi/2$ phase
shift. In the Anderson model, both spin and charge degrees of freedom
are present. 
 When only exchange-type interactions are considered in the context of the
 coarse-grained Hamiltonian, there are
two types of such interactions: $J_s{\bf S}\cdot {\bf J}(0)$ and $J_c{\bf
I}\cdot {\bf {I}}(0)$. Here ${\bf J}(0)$ and ${\bf I}(0)$ represent
the spin and axial charge current for conduction electrons, respectively. 
For large $U$, only the sector with ${\bf I}^2=0$ is relevant. Then, the spin-spin exchange
coupling describes the low-energy physics, and thus the model reduced to the Kondo model.  
For small
$U$, the charge-charge exchange interaction $J_c$ becomes compatible 
with the spin-spin exchange interaction $J_s$. 

Now, 
one can realize that there are 
similarities between the Anderson model and the present one; $J_s$ corresponds to $J$ in Eq. (\ref{Heff2_S})
and $J_c$ to $K$ and ${\bf I}(0)({\bf I})$ to ${\bf L}(0)({\bf L}_I)$ in
Eq. (\ref{Heff2}). As for the fusion process, the axial charge fusion,
{\it i.e.,} the $\pi/2$ phase shift, in
the Anderson model corresponds to the SO(5)-${\bf 4}$ fusion in
the present model.

The spin-1/2 and SO(5)-${\bf 4}$ fusions introduced above are equivalent in the sense that the
spin-1/2 fusion and the $\pi/2$ phase shift are equivalent in the Kondo or
Anderson model. Our answer to the question ``{\it what is going on for small
$U$?}'', which is the main motivation in this paper, is that the NFL
spectra of 2CKM can be obtained via the
nonmagnetic SO(5)-${\bf 4}$ fusion, and thus, we can interpret the
low-energy physics for small $U$ as 
 the nonmagnetic SO(5) Kondo effects in the same way that the
 conventional Kondo effects can be interpreted as the strong potential
 scattering with the phase shift $\pi/2$. 
Of course, one can still interpret it  
 as a magnetic one but the SO(5)-$\bf 4$ fusion is much better for 
 understanding the physics for small $U$, 
since what makes the low-energy physics for small $U$
different from that for the Kondo regime is the nonmagnetic degrees of
freedom. 
Indeed, as will be shown in
Sec. \ref{sec-NRG}, the residual interactions around the NFL fixed point
for small $U$ 
are governed by the operators in the SO(5) sector rather than those in
the spin sector.

\subsubsection{Leading irrelevant operators} \label{irreOp}
Low-temperature thermodynamic properties are governed
by leading irrelevant operators around the fixed point. In the Kondo
regime, {\it i.e.,} for large $U$, the
leading irrelevant operator should be in the spin sector, and it is ${\bf J}_{-1}\cdot\vec{\phi}_s$, with
${\vec \phi}_s$ being spin-SU(2) primary fields with the dimension
$\Delta=1/2$ and the quantum numbers $(j,{\rm SO(5)})=(1,{\bf 1})$ in
Table \ref{tbl-1} (c).
In total, the dimension of this
operator is $3/2$. In the presence of this operator, it is
well known\cite{Aff2,Aff3} that the impurity contribution to the 
specific heat $C$ is proportional to $-T\ln T$, and
the magnetic susceptibility $\chi_s$ diverges logarithmically $\chi_s\sim -\ln
T$ at low temperatures.

For small $U$, we expect that the operator in the spin sector does not
play an important role, and thus, operators in the SO(5) sector dominate. 
Then, the situation is analogous to 
the spin-3/2 dipole-octupole Kondo model.\cite{hat1} Since 
the first descendants of primary fields $\vec{\phi}_{\bf 5}$ with $(j,{\rm
SO(5)})=(0,{\bf 5})$ and $\Delta=1/2$ in Table \ref{tbl-1} (c), ${\bf L}_{-1}\vec{\phi}_{\bf 5}$, cannot form 
an SO(5) singlet, the leading irrelevant operator is the energy-momentum tensor
in the SO(5) sector at the impurity site :${\bf L}(0)\cdot{\bf L}(0)$:. The leading
 dimension of this operator is $2$, {\it i.e.}, the ``Fermi liquid'' like
 interaction.\cite{Aff1} This readily
 indicates that the low-temperature impurity specific heat $C\propto T$.
As investigated in Ref. 34, the susceptibilities of the SO(5)
vector ${\bf 5}$, $\vec{\phi}_{\bf 5}$, 
diverge logarithmically $\sim -\ln T$, indicating the divergence of the
susceptibility of ${\bf n}_I$. The susceptibility of the SO(5)
generators is independent of $T$ at low temperatures, since the
dimension of the 
$(0,{\bf 10})$ operator in Table \ref{tbl-1} (c) is $\Delta=1$. Thus, the susceptibilities of
${\bf L}_I$ are Fermi liquid like. We call these behaviors SO(5) NFL hereafter.

Here, we notice that the SO(5)
vector in our model corresponds to $\tau_yI_{\alpha}$ and $\tau_{\beta}\mathcal P$
with $\alpha=x,y,\ {\rm or}\ z$ and $\beta=x,\ {\rm or}\ z$, see Table
\ref{tbl-0} and Eq. (\ref{nI}). Physically, these operators 
correspond to bipolaron fluctuations, which are coupled fluctuations of
the flavor and the axial charge. In the original variables $b$ and
$b^{\dagger}$, $\tau_x \pm i\tau_y$ are, roughly speaking, related to $b$
and $b^{\dagger}$.


In principle, there exist both terms 
${\bf J}_{-1}\cdot \vec{\phi_s}$ and ${\bf L}(0)\cdot{\bf L}(0)$ 
for general values
of $U$, since the original interaction is in complex form of the spin
and SO(5) degrees of freedom and also terms that cannot be described by
exchange forms. What varies as a function of parameters, 
{\it e.g.}, $U$ along the critical line, is the coefficients
of these operators in the effective Hamiltonian near the fixed point. 
Such a situation is represented by the following residual effective
Hamiltonian:
\begin{eqnarray}
\delta H_{{\rm eff}}=\lambda_s{\bf J}_{-1}\cdot\vec{\phi}_s
+\lambda_L{\bf L}_{-1}\cdot {\bf L}_{-1},\label{delH}
\end{eqnarray}
where $\lambda_s$ and $\lambda_L$ depend on microscopic parameters such
as $U$, $V_0$, and $V_1$. We have retained the leading term of ${\bf
L}(0)\cdot {\bf L}(0)$ in the second
term in Eq. (\ref{delH}) and we have not included a term ${\bf
J}_{-1}\cdot {\bf J}_{-1}$, which has the dimension 2, 
since it is sufficient to include only the 
 leading irrelevant
operators in each of the sectors in the following analysis.
As we investigated above, the relative magnitude of the two terms varies, 
and the first term is dominant in the Kondo regime, while
the second one prevails in the weak coupling regime. 

An interesting crossover is expected especially in the impurity
contributions to  specific
heat $C$. Since two sectors are decoupled, $C$ is the sum of the 
contribution of each sector:\cite{Aff1,Aff3}
\begin{eqnarray}
C=-\gamma_sT\ln\Big(\frac{T}{T_s}\Big)+\gamma_LT. \label{speci}
\end{eqnarray}
Here, $T_s$ is a dynamically generated energy scale in the spin sector,
which is proportional to the Kondo temperature in the Kondo regime. The
parameter $\gamma_s(\gamma_L)$ is proportional to
$\lambda_s^2(\lambda_L)$.\cite{Aff1,Aff3} 
As analyzed by Johannesson {\it et al.},\cite{2cam} $\lambda_s^2\sim 1/T_s$ and
$\lambda_L\sim 1/T_L$, where $T_L$ is the ``Kondo temperature'' for the
SO(5) sector.
The crossover temperature  $T^*$ can be defined by the temperature where
the magnitudes of the 
two terms in Eq. (\ref{speci}) are equal, and is given by
\begin{eqnarray}
T^*=T_s\exp\Big(-\frac{\gamma_L}{\gamma_s}\Big).\label{Tstar}
\end{eqnarray}
For $T<T^*$, the specific heat due to the spin sector dominates, and thus,
$C/T$ diverges logarithmically. However, for sufficiently small $U$,
$\gamma_L/\gamma_s\sim T_s/T_L\gg 1$, thus,  $T^*$ is never reached in a
realistic
temperature range and $C/T$ stays constant at low temperatures.
This is a marked contrast between the NFL in the 2CKM and the SO(5) NFL.

Finally, let us comment on the ``secondary-diverging'' susceptibility in
each of the parameter regime. Even for large $U$, the susceptibility of
$\vec{\phi}_{\bf 5}$, indeed, diverges logarithmically. 
We call this divergence "secondary", since the coefficient of this 
part is expected to be very small for large $U$:\cite{2cam} $\sim -[\ln(T/T_L)]/T_L$ with $T_L\sim U/2$.
The same is
true for the spin susceptibility for small $U$. There, the system is in
the valence fluctuation regime and the spin
susceptibility is $\sim -[\ln(T/T_s)]/T_s$ with $T_s$ proportional to the
hybridization width.

\subsubsection{Stability of the fixed points} \label{stability}
In this subsection, we investigate the stability of the NFL
fixed point derived in Sec. \ref{irreOp} against various perturbations.

First, we investigate symmetry breaking fields. In Table \ref{tbl-1} (c), 
there are SO(5)-${\bf 5}$ primary fields $\vec{\phi}_{\bf 5}$
 with the dimension $\Delta=1/2$. Thus, when the SO(5) symmetry is broken,
 a term $\vec{h}_{\bf 5}\cdot \vec{\phi}_{{\bf 5}}$ appears 
in the effective Hamiltonian, which is relevant, and
thus the NFL fixed point is unstable against this perturbation.
Practically speaking, the SO(5) symmetry-breaking field $\vec{h}_{\bf
5}$ includes the inversion 
symmetry-breaking field and the flavor (even-odd) symmetry-breaking one. 

In the presence of the 
inversion symmetry breaking field,
$(\vec{\phi}_{\bf 5})_2$ with the quantum number $(j,i,P,F)=(0,0,1,0)$ appears in
the effective Hamiltonian. Here $(\vec{\phi}_{\bf 5})_i$ represents the
$i$th field in the five component vector $\vec{\phi}_{\bf 5}$.
With regard to the flavor symmetry breaking field, $(\vec{\phi}_{\bf 5})_4$ 
with $(j,i,P,F)=(0,0,0,1)$ appears in the effective Hamiltonian. The
Hamiltonian (\ref{H}) has inversion symmetry, while the flavor symmetry
is higher than that of the original Hamiltonian (\ref{H}) and is
realized only along the critical line. Thus, $(\vec{\phi}_{\bf 5})_2$
cannot appear even away from the critical line, 
while $(\vec{\phi}_{\bf 5})_4$ can appear away from the critical line and $(\vec{h}_{\bf 5})_4(\vec{\phi}_{\bf 5})_4$ 
is the perturbation that makes the NFL fixed points unstable. 
This flavor
symmetry breaking causes the energy difference between $|\!\!\Uparrow\rangle$
and $|\!\!\Downarrow\rangle$. 
 The flavor symmetry breaking is, indeed, consistent with
the spectra in the NRG and in the
small clusters as analyzed in Sec. \ref{SO5local}.

Another relevant field is the magnetic field $h$, since there is 
 an SO(5)-singlet and spin-1 primary fields $\vec{\phi}_s$ in Table
\ref{tbl-1} (c) with $\Delta=1/2$, which couple with $h$. This is the
 same as in the 2CKM\cite{Aff4} and we do not discuss it in detail.

Finally, particle-hole asymmetry is marginal,
since the operator with $(j,i_{i_z},P,F)=(0,1_0,0,0)$ is classified in $[j,{\rm
SO(5)}]=(0,{\bf 10})$ and the dimension is $\Delta=1$ in Table
\ref{tbl-1} (c). This operator breaks the SO(5) symmetry. It is well known that  
the potential scattering $VI_z(0)$ is 
absorbed in phase shifts\cite{Cox}, 
and thus, the NFL properties are not affected
 except for the specific heat for small $U$, as we discuss below. 
When particle-hole symmetry is broken
 by charge-conserved perturbations such as $VI_z(0)$, in
 addition to (\ref{delH}) with anisotropic exchange interactions in the
 SO(5) sector [see, Eq. (\ref{SO5aniso})], 
 anisotropic flavor exchange interactions are allowed to appear, leading to
 the additional residual interactions, 
\begin{eqnarray}
\delta H_{\rm eff}^{f}\sim g_xL_{-1}^{34}(\vec{\phi}_{\bf 5})_2+ 
g_yL_{-1}^{24}(\vec{\phi}_{\bf 5})_3+ g_zL_{-1}^{23}(\vec{\phi}_{\bf 5})_4.
\label{delHfl}
\end{eqnarray}
Here, $(-L^{34}_{-1},L^{24}_{-1},L^{23}_{-1})={\bf F}_{-1}$ is the
flavor current operator defined in
Sec. \ref{SO5SU2} and $g_i(i=x,y,\ {\rm or}\ z)$ is constant
proportional to the symmetry breaking field $\sim V$. 
Note that this form is not SO(5) invariant. However, it is still invariant under
the inversion and the spin 
SU(2) operations, and also the total charge is conserved.
 Since the scaling dimension
of (\ref{delHfl}) is $\Delta=3/2$, the impurity contribution of the
specific heat of this
term is similar to that of the spin sector in Eq. (\ref{speci}).

Second, we investigate exchange anisotropies in the spin and the SO(5)
sectors. 
The irrelevance of the spin exchange anisotropy is
explained in the same way as in the case of the magnetic
2CKM.\cite{Aff4}
As for the exchange anisotropy in the SO(5) sector, 
when the anisotropy exists, the isotropic effective interaction $K{\bf L}_I(0)\cdot {\bf L}(0)$ is replaced,
\begin{eqnarray}
K{\bf L}_I\cdot {\bf L}(0)&\to&K{\bf L}_I\cdot {\bf L}(0)
+\delta K'\tau_z{\bf I}\cdot {\bf I}'(0)\nonumber\\
&&-\delta K''\tau_x{\bf I}\cdot{\bf I}''(0)
-\frac{1}{2}\delta K'''\tau_y{\mathcal P\mathcal D}(0), \ \ \ \ \label{SO5aniso}
\end{eqnarray}
where $\delta K'$, $\delta K''$ and $\delta K'''$ are the deviations
 from the isotropic interactions. Note that Eq. (\ref{SO5aniso}) is
 charge-SU(2) invariant.
 This anisotropy generates residual interactions such as $L^{ab}_{-1}L^{ab}_{-1}$ with some sets of $ab$. This operator has 
dimension $2$, {\it i.e.}, it is irrelevant.

\section{Numerical Renormalization Group Results} \label{sec-NRG}
In this section, we examine the crossover from the SO(5) NFL to the NFL in
the magnetic 2CKM as $U$ increases along the critical line by using 
Wilson's NRG.\cite{nrg} One of the advantage in using
NRG is that we can obtain information about the scaling dimensions
$\Delta$ of
leading irrelevant operators around fixed points by analyzing spectra
obtained in NRG.\cite{zarand,koga} The details of the NRG method are
explained in a 
previous paper.\cite{Yashiki2} Here, we will analyze variations of 
NRG spectra along the critical line. A detailed analysis of physical
quantities will be reported elsewhere.\cite{UnpublishHat}

In this section, we will show data for three
different parameters: large $U=0.8\tilde{D}$, small $U=0.02\tilde{D}$,
and intermediate $U=0.2\tilde{D}$. 
 Here, $\tilde{D}$ is related to half of the band
 width $D$ of conduction
 electrons for both $s$ and $p$ bands as $\tilde{D}=D(\Lambda+1)/(2\Lambda)$, and the Fermi energy is at the
 middle of the band. Here, $\Lambda$ is a discretization parameter in NRG
 and we use $\Lambda=3$. For each value of $U$, we tune the hybridization $V_1$ to
 realize the NFL fixed point, while $V_0=0.2\tilde{D}$ and 
 $\Omega=0.2\tilde{D}$ are fixed. The resulting $V_1$'s are $V_1\simeq 
 0.16181\tilde{D}$ for $U=0.02\tilde{D}$, $V_1\simeq 0.15623\tilde{D}$ for
 $U=0.2\tilde{D}$, and $V_1\simeq 0.15473\tilde{D}$ for $U=0.8\tilde{D}$.

In the calculations, we utilize the spin rotational symmetry to restore
 the states and 5000 states labeled by the set of quantum numbers
 $(j,i_z,P)$ are kept at each iteration in NRG. As for the number of
 local phonon states, we use 20 phonon states in our calculations.

\subsection{NRG spectra}
\begin{figure}[t]
\begin{center}
    \includegraphics[width=0.5\textwidth]{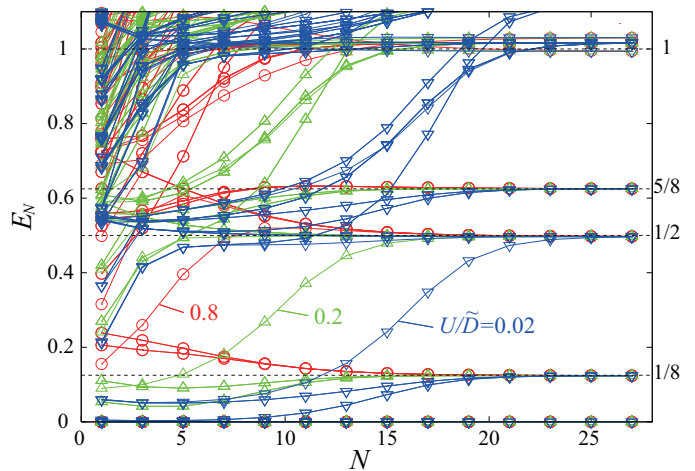}
\end{center}
\caption{(Color online) NRG spectra as a function of the renormalization
 group step $N$ with $N$ odd for $U=0.8\tilde{D}$($\bigcirc$),
 $U=0.2\tilde{D}$($\bigtriangleup$) and
 $U=0.02\tilde{D}$($\bigtriangledown$). 
The spectra are uniformly scaled such that the first excited energy
 becomes $1/8$ for large $N$, and the identical 
factor is used for all three.
For all three, the spectra
 converge on the NFL spectra $E_{\rm NFL}$ (dotted lines) 
for large $N$. 
 The small deviations from $E_{\rm NFL}$ are due
 to truncation errors in the NRG calculations.
 }
\label{fig-1}
\end{figure}

Let us show the NRG spectra $E_N$ as a function of the renormalization group step
$N$ in Fig. \ref{fig-1}. Apart from differences in crossover scale
$N_0$($\sim$$15$, $9$, and $3$ for $U/\tilde{D}=0.02$, $0.2$, and $0.8$,
respectively), all three spectra converge on the
NFL spectra $E_{\rm NFL}$ shown in Table \ref{tbl-1} (b). Down to the
lowest energy scale, we also confirm that the impurity contribution to
the entropy is $\ln\sqrt{2}$, as expected in the 2CKM\cite{Aff-ent}
(see, Fig. \ref{fig-specificheat}). 
For the smallest $U$, the
crossover step $N_0$ is large and this is due to the fact that 
for $N<N_0$, the  nonmagnetic ``local moment'' fixed point is realized.\cite{Yashiki2,nonmagfix}

In Table \ref{tbl-nrg}, the low-energy states at the NFL fixed point 
and their energy with
eigenvalues of the z-component of the total axial charge, the total
spin, and the total parity for $U/\tilde{D}=0.02$ are listed. 
For other two $U$'s, the results are very similar. NRG energy
eigenvalues and their quantum number are consistent with those derived by
the BCFT.
The energy
spectra for the odd-$N$ NRG
step and even-$N$ one are identical within the numerical accuracy 
except for the parity eigenvalue $P$. The eigenvalue $P$ for the
even-$N$ 
step is obtained from that for the odd-$N$ one simply by interchanging even ($P=0$) and odd ($P=1$)
for all the states. Here, the ground state of the free electron system 
for the odd (even) $N$ 
is non-degenerate (degenerate).
This is easily understood by noting that the
free spectra for even $N$ can be obtained by shifting the conduction
electron charge $Q_{s(p)}$ to $Q_{s(p)}+1$,\cite{Cox} where $Q_{s(p)}$ 
is the charge for $s(p)$-wave conduction electrons. As a result, the
total parity $P={\rm mod}(Q_p+b^{\dagger}b,2)$ is shifted to $\to 
{\rm mod}(Q_p+1+b^{\dagger}b,2)={\rm mod}(P+1,2)$. Thus, the even
(odd) parity states are simply relabeled as odd (even) parity states
and then the fusion process leads to the NFL spectra with $P$ replaced
by mod$(P+1,2)$. 

\begin{table}[!t]
\caption{Spectra $E_N$ at the NFL fixed point for odd $N=29$ (even $N=30$) and
 $U=0.02\tilde{D}$. 
 The
 parity $P$ and the NRG spectra $E_N$ for even $N$ are shown in
 parentheses. $E_N$ is scaled such that the first excited energy
 with * 
 becomes $1/8$ in order to compare $E_N$ with the spectra
 obtained by the BCFT $E_{\rm BCFT}$. 
}
   \begin{tabular}{cccccc}
\hline
\hline
\ \ \ $j$ \ \ \ & \ \ \ $i_z$ \ \ \          & $P$    & SO(5)      &
    $E_{\rm BCFT}$ & $E_{29}$ $(E_{30})$ \\
\hline
$\frac{1}{2}$ & $0$             & $0(1)$  & ${\bf 1}$ & $0$           &
			$0\ \  (0)$ \\
\hline
$0$           & $-\frac{1}{2}$  & $0(1)$  & ${\bf 4}$ & $\frac{1}{8}$ &
			$0.12503\ (0.125^*)$ \\
$0$           & $-\frac{1}{2}$  & $1(0)$  &           &               &
			$0.12500\ (0.12504)$ \\
$0$           & $\frac{1}{2}$   & $0(1)$  &           &               &
			$0.12503\ (0.12500)$ \\
$0$           & $\frac{1}{2}$   & $1(0)$  &           &               &
			$0.125^*\ \ (0.12504)$ \\
\hline
$\frac{1}{2}$ & $-1$            & $1(0)$  & ${\bf 5}$ & $\frac{1}{2}$ &
			$0.50155\ (0.50154)$ \\
$\frac{1}{2}$ & $0$             & $0(1)$  &           &               &
			$0.50154\ (0.50153)$ \\
$\frac{1}{2}$ & $0$             & $1(0)$  &           &               &
			$0.50124\ (0.50136)$ \\
$\frac{1}{2}$ & $0$             & $1(0)$  &           &               &
			$0.50155\ (0.50154)$ \\
$\frac{1}{2}$ & $1$             & $1(0)$  &           &               &
			$0.50155\ (0.50154)$ \\
\hline
$1$           & $-\frac{1}{2}$  & $0(1)$  & ${\bf 4}$ & $\frac{5}{8}$ &
			$0.63098\ (0.63104)$ \\
$1$           & $-\frac{1}{2}$  & $1(0)$  &           &               &
			$0.63100\ (0.63100)$ \\
$1$           & $\frac{1}{2}$   & $0(1)$  &           &               &
			$0.63098\ (0.63104)$ \\
$1$           & $\frac{1}{2}$   & $1(0)$  &           &               &
			$0.63100\ (0.63100)$ \\
\hline
$\frac{3}{2}$ & $0$             & $0(1)$  & ${\bf 1}$ & $1$           &
			$1.00306\ (1.00304)$ \\
\hline
$\frac{1}{2}$ & $-1$             & $0(1)$  & ${\bf 10}$ & $1$
		    & $1.02581\ (1.02600)$ \\
$\frac{1}{2}$ & $-1$             & $0(1)$  &            &
		    & $1.02616\ (1.02620)$ \\
$\frac{1}{2}$ & $-1$             & $1(0)$  &            &
		    & $1.02615\ (1.02619)$ \\
$\frac{1}{2}$ & $0$              & $0(1)$  &            &
		    & $1.02581\ (1.02600)$ \\
$\frac{1}{2}$ & $0$              & $0(1)$  &            &
		    & $1.02616\ (1.02620)$ \\
$\frac{1}{2}$ & $0$              & $1(0)$  &            &
		    & $1.02615\ (1.02619)$ \\
$\frac{1}{2}$ & $0$              & $1(0)$  &            &
		    & $1.02580\ (1.02599)$ \\
$\frac{1}{2}$ & $1$              & $0(1)$  &            &
		    & $1.02581\ (1.02600)$ \\
$\frac{1}{2}$ & $1$              & $0(1)$  &            &
		    & $1.02616\ (1.02620)$ \\
$\frac{1}{2}$ & $1$              & $1(0)$  &            &   	    & $1.02615\ (1.02619)$ \\
\hline
$\frac{1}{2}$ & $0$              & $0(1)$  & ${\bf 1}$  & $1$	    & $1.03984\ (1.03992)$ \\
\hline
\hline
   \end{tabular}\label{tbl-nrg}
\end{table}

The point we address in the following is how $E_N$ varies as a function of $N$ for
the three parameters. Near the fixed point, the spectrum at the step $N$, 
$E_N$, is represented as 
\begin{eqnarray}
E_N=E_{\rm NFL}+\delta E_N,
\end{eqnarray}
and the deviation from the fixed point value $\delta E_N$ is given
by\cite{nrg, zarand, koga}
\begin{eqnarray}
\delta E_N=\sum_r \lambda_r \Lambda^{-\frac{(\Delta_r-1) N}{2}}.\label{delEN}
\end{eqnarray}
Here, $r$ identifies leading irrelevant operators appearing near the
fixed point. For the single channel Kondo model, $\Delta_r=2$.\cite{nrg} In magnetic
2CKM, $\Delta_r=3/2$,\cite{koga} which represents
the ``slower'' renormalization than in the single channel model. 
Note that $\Delta_r$ coincides with the scaling dimension of the operator.
In the following, we analyze $\delta E_N$ in details and examine the crossover
predicted in Sec. \ref{BCFT}.

\begin{figure}[t!]
\begin{center}
   \includegraphics[width=0.5\textwidth]{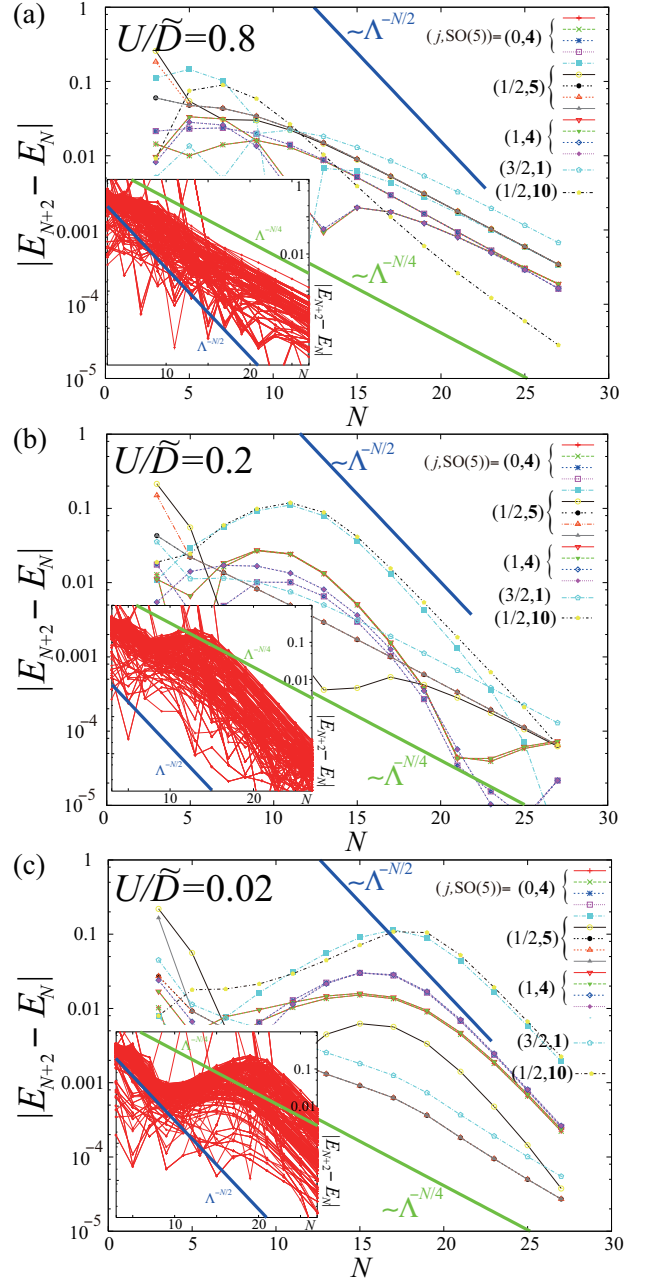}
\end{center}
\caption{(Color online) $|E_{N+2}-E_{N}|$ for 15 low-energy states 
as a function of the renormalization
 group step $N$ with $N$ odd for (a) $U=0.8\tilde{D}$, 
 (b) $U=0.2\tilde{D}$, and
 (c) $U=0.02\tilde{D}$. Each state is labeled by the spin $j$ and the
 dimension of the irreducible representation in the SO(5) group. 
A straight line  $\sim \Lambda^{-N/2}$ is characteristic of the
 irrelevant operator in the SO(5) sector ${\bf
 L}_{-1}\cdot {\bf L}_{-1}$, while $\sim \Lambda^{-N/4}$ 
is characteristic of the operator in the spin sector ${\bf J}_{-1}\cdot \vec{\phi}_s$.
Inset: the same plot for
 300 low-energy states.
 }
\label{fig-2}
\end{figure}

\subsection{Crossover along the critical line}
The deviation from the fixed point $\delta E_N$ contains information about
leading irrelevant operators as shown in Eq. (\ref{delEN}). 
In order to evaluate $\delta E_N$, we use $E_{N+2}-E_{N}$:
\begin{eqnarray}
E_{N+2}-E_{N}=\sum_r\lambda_r 
\Lambda^{-\frac{(\Delta_r-1) N}{2}}(\Lambda^{1-\Delta_r}-1). \label{subtract}
\end{eqnarray}
Thus, when there is a dominant leading irrelevant term with $\Delta_r=\Delta$, $E_{N+2}-E_{N}$
is proportional to $\Lambda^{-(\Delta-1)N/2}\sim \delta E_N$. Here, we use $(N+2)$- and
$N$-step eigenvalues in Eq. (\ref{subtract}), since, in NRG, there is
even-odd alternation in the spectra. 

Figure \ref{fig-2} shows $|E_{N+2}-E_{N}|$ of the low-energy states 
for the three parameters. Each state is labeled by the spin and the SO(5)
indices: $(j,{\rm SO(5)})$.  
For the largest $U=0.8\tilde{D}$, 
it is clear that the scaling dimension $\Delta$ is $\Delta=3/2$, and
thus, the NFL is described by the 
magnetic 2CKM as investigated in Sec. \ref{BCFT}. 
For the smallest $U=0.02\tilde{D}$, the dimension of leading irrelevant
operator is $\Delta=2$, since for most of the states the $N$ dependence
is $\sim \Lambda^{-N/2}$. This is consistent with our analysis in
Sec. \ref{irreOp}. 
The $\Lambda^{-N/2}$ dependence is due to the existence of
the nonmagnetic SO(5) residual interaction ${\bf L}_{-1}\cdot {\bf
L}_{-1}$. In principle, there is a possibility that the term ${\bf
J}_{-1}\cdot {\bf J}_{-1}$ is the origin of the $\Lambda^{-N/2}$
dependence. This possibility, however, is unlikely from a physical
standpoint. It is unphysical that only the coefficient of the leading term
in the spin sector is suppressed, while that of the sub-leading term in
the same sector is not, as $U$ decreases.

One may find that some of the curves follow the $\Lambda^{-N/4}$
dependence for $U=0.02\tilde{D}$, but
the absolute value is very small, i.e., $|\lambda_{\frac{3}{2}}|$ is
very small, where we use $\Delta_r$ as the index $r$. 
Although, in principle, there exist contributions of matrix elements of
the operators in the effective Hamiltonian to
$|E_{N+2}-E_N|$, it is unlikely that the small absolute value is only
due to the matrix elements, and thus, we neglect them in the
following analysis. 
This $N$
dependence, $\Lambda^{-N/4}$, must originate in the 
 magnetic interactions ${\bf J}_{-1}\cdot \vec{\phi}_s$, since in the
 SO(5) sector there are no such operators that generate this $N$ dependence.
 Note that in NRG the step $N$ is related
to the energy scale as $D\Lambda^{-N/2}$, and, for example, $N=20$
corresponds to $3^{-10}D\simeq 1.69\times 10^{-5}D$. At $N=20$, the
absolute value of $|\lambda_2|$ is more
than 100 times larger than $|\lambda_{\frac{3}{2}}|$.
Thus, we expect that 
there the spin degrees of freedom have no effect on, for example,
the specific heat since $|\lambda_2|\gg |\lambda_{\frac{3}{2}}|$. 
As for the spin susceptibility, 
a logarithmic divergence with a very small coefficient is expected,
reflecting the small $\lambda_{\frac{3}{2}}$. This is similar to
the case of the flavor susceptibility in the two-channel Anderson model.\cite{2cam}

As expected, the situation changes as $U$ increases. For
$U=0.2\tilde{D}$, it is clear that both $\lambda_{2}$ and
$\lambda_{\frac{3}{2}}$ are present with similar magnitudes. 
Around $N\sim 22$, the crossover from the SO(5) NFL to the NFL in 2CKM
occurs. Thus, from our NRG calculations, it is clear that 
the profile of the leading irrelevant
operators changes smoothly from the weak-coupling regime to the Kondo
regime. 
These results confirm the results in Sec. \ref{BCFT}.

\begin{figure}[t]
\begin{center}
   \includegraphics[width=0.5\textwidth]{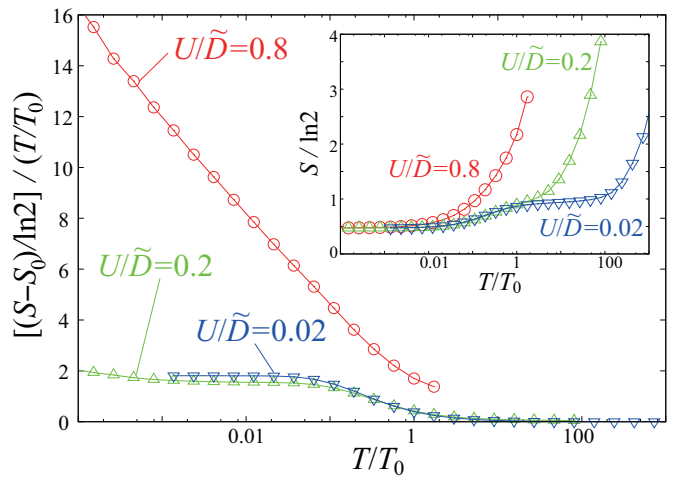}
\end{center}
\caption{(Color online) Temperature dependence of $(S-S_0)/T$ for the three
 values of $U$. Temperature $T$ is scaled by $T_0$; $T_0/{\tilde{D}}=0.192$ for $U/{\tilde{D}}=0.8$,
 $T_0/{\tilde{D}}=0.00713$ for $U/{\tilde{D}}=0.2$, and
 $T_0/{\tilde{D}}=0.000264$ for $U/{\tilde{D}}=0.02$. 
 Inset:  Temperature dependence of $S$.}
\label{fig-specificheat}
\end{figure}

Finally, we discuss the impurity contributions of specific heat $C$ and the
impurity entropy $S$. Figure \ref{fig-specificheat} shows the temperature
dependence of $(S-S_0)/T$ and $S$ for the three parameters of $U$ with
$S_0\simeq \ln\sqrt{2}$. Since $(S-S_0)/T\simeq C/T$ 
at low temperatures, it represents $C/T$ for $T<T_0$. 
Crossover
temperatures are defined as $T_0\equiv \tilde{D}\Lambda^{-N_0/2}$.   
For large
$U/{\tilde{D}}=0.8$, $C/T$ at low temperatures diverges logarithmically and this is consistent with the
conventional magnetic 2CKM. As expected from the results of the scaling
dimension of leading irrelevant operators, the temperature dependence of
$C/T$ changes with decreasing $U$. One can clearly see that it is
constant at low temperatures for $U/\tilde{D}=0.02$ and $0.2$. 
For $T<T_0$, the
temperature dependence of entropy for $U/\tilde{D}=0.02$ and $0.2$ is  
well described by a single-scale function of $(T/T_0)\sim (T/T_L)$, while that for
$U/\tilde{D}=0.8$ has a different functional form, since $T_0\sim T_s$ for
$U/\tilde{D}=0.8$. 
For
$U/\tilde{D}=0.2$, the temperature dependence changes from $\sim$
const. to $-\ln T$, and the crossover temperature has been defined as
$T^*$ in Eq. (\ref{Tstar}). This is the crossover from the SO(5)-operator dominant regime
to the spin-operator dominant regime. Below $T\sim T^*$, the
temperature dependence of $S$ is not described by the single-scale
function of $(T/T_L)$, although the deviation in $S$ is very small $\sim
-T\ln T$.
Note that the impurity
entropy $S$ is $S\simeq \ln \sqrt{2}$ for the temperature  
where $C/T=$ const. appears. 
These results confirm our BCFT
predictions and the importance of nonmagnetic SO(5) degrees of freedom
for small $U$.

\section{Discussion and Summary}

In this paper, we have analyzed low-energy critical theory in a
two-channel Anderson model with phonon-assisted hybridization on the
basis of BCFT and NRG. One important finding is that nonmagnetic SO(5)
degrees of freedom are constructed in the ${\bf I} \otimes \vec{\tau}$
sector, which are ``hidden'' in the 2CKM due to the Hilbert space 
restriction, and also the conduction electron part is rewritten by the
SO(5) currents. These nonmagnetic degrees of freedom are important for
small $U$.
 We have demonstrated that
SO(5)-${\bf 4}$ fusion gives exactly the same NFL spectra
as those in the magnetic 2CKM. The difference between the spin-1/2 
and
the SO(5)-${\bf 4}$ fusions have been discussed and we interpret the fusion as
simply equivalent ones, noticing that the spin-1/2 fusion and the $\pi/2$ phase shift in the
Anderson model are equivalent. A full understanding of the exact fusion
process will require a 
more sophisticated analysis. 

In the form of residual effective
interaction (\ref{delH}), the crossover between small and large $U$ can
be described by changes in the coefficients $\lambda_s$ and $\lambda_L$
in Eq. (\ref{delH}).  Note that the residual interactions in the SO(5) sector
never appear if we map the model
to the magnetic 2CKM, since there is no  $\lambda_L{\bf L}(0)\cdot {\bf L}(0)$
term in the perturbation expansion in the spin sector. 
Using physical intuition, we predict that the SO(5) sector is more
important than the spin sector for small $U$, leading to 
linear specific heat at low temperatures. This has been checked by
the NRG calculations; the scaling dimension of leading irrelevant
operators varies from $\Delta=3/2$ to $\Delta=2$ as $U$ decreases, and 
 the impurity contribution of specific heat is proportional to
 temperature for small $U$. 

The difference between the present NFL for small $U$ and the NFL in the two-level systems
should be noted, although they both have a nonmagnetic origin. It is well known that the 
NFL spectra in the two-level Kondo model is derived via flavor-1/2
fusion.\cite{Cox} The spectra is the same as those in the magnetic 2CKM when the
spin and the flavor sector are interchanged, while the present NFL
spectra are 
exactly the same as those in the magnetic 2CKM. The scaling dimension of 
 the leading irrelevant operator in the two-level Kondo model is 3/2, which
leads to a logarithmic diverging specific heat coefficient. Thus,
nonmagnetic Kondo phenomena in the two models should be distinguished
and the
microscopic mechanism for the NFL in each of the models is quite
different, i.e., flavor SU(2) and nonmagnetic SO(5) exchange interactions.

Our BCFT analysis can also be applicable to the anharmonic model investigated in
Ref. 24, since even in the anharmonic phonon model, parity $P$ is a good
quantum number and the phonon states in the effective
theory would be described by $\vec{\tau}$ as in a similar manner to that
of the 
present analysis. 
With regard to the generalization of this model to the more realistic one, 
it is interesting to take into account optical
modes that couple with localized electrons. This
electron-phonon coupling reduces the bare Coulomb repulsion. When it 
is sufficiently large, the effective Coulomb
interaction becomes attractive, and thus, it is possible to realize
another NFL fixed point in which the spin and charge sector are 
interchanged from the present NFL for $U>0$.
With regard to a lattice generalization of the present model, it is also interesting
to examine whether some composite pairing operators listed in Table {\ref{tbl-0}}
condensate into exotic superconducting states.

In summary, we have investigated the microscopic origin of the line of non-Fermi liquid
fixed points found previously by numerical
simulations.\cite{Daggoto,Yashiki1,Yashiki2} 
We have succeeded in constructing nonmagnetic SO(5) degrees of freedom, 
and, on the basis of
boundary conformal field theory, we have pointed out that, for the weak
electron-electron correlation regime, the non-Fermi liquid can be
interpreted as an SO(5) non-Fermi liquid, which crosses overs to a non-Fermi liquid in the Kondo regime. 
We have also analyzed the
difference in the leading irrelevant operators as $U$ varies, and indeed
we have confirmed it by numerical
simulations. In particular, the impurity contributions to the specific
heat 
are proportional to temperature $T$ for small $U$, while they are proportional to
$-T\ln T$ for large $U$. The present results demonstrate that it is
important to take into account not only single degrees of freedom, {\it e.g.},
only a phonon, but also 
complex degrees of freedom formed both by electrons and phonons in the Kondo
problems in electron-phonon coupled systems.

\section*{Acknowledgment}
The author thanks K. Ueda, T. Hotta, H. Tsunetsugu and S. Yashiki 
for fruitful discussions.
This work was supported by a Grant-in-Aid for Scientific Research on Innovative Areas "Heavy Electrons" (No. 23102707) of The Ministry of Education, Culture, Sports, Science, and Technology, Japan.

\appendix
\section{Matrix representations of SO(5) Generators and Vectors}
In this Appendix, we summarize the definitions of SO(5) matrices. We follow
the convention used by Wu, {\it et al.}\cite{Wu}
\subsection{Generators:\ {\bf 10} representation}\label{SO}
The SO(5) generators $\mathcal L^{ab}$ define all the SO(5) rotation 
and are given by ${\mathcal L}^{ab}\equiv \Gamma^{ab}/2$ with 
\begin{eqnarray}
\Gamma^{12}&=&-
\begin{pmatrix}
0&0&0&1\\
0&0&1&0\\
0&1&0&0\\
1&0&0&0
\end{pmatrix}, \ \ 
\Gamma^{13}=
\begin{pmatrix}
0&0&0&i\\
0&0&-i&0\\
0&i&0&0\\
-i&0&0&0
\end{pmatrix},\nonumber\\
\Gamma^{14}&=&
\begin{pmatrix}
0&0&-1&0\\
0&0&0&1\\
-1&0&0&0\\
0&1&0&0
\end{pmatrix}, \ \ 
\Gamma^{15}=
\begin{pmatrix}
-1&0&0&0\\
0&-1&0&0\\
0&0&1&0\\
0&0&0&1
\end{pmatrix},\nonumber\\
\Gamma^{23}&=&
\begin{pmatrix}
1&0&0&0\\
0&-1&0&0\\
0&0&1&0\\
0&0&0&-1
\end{pmatrix}, \ \ 
\Gamma^{24}=
\begin{pmatrix}
0&i&0&0\\
-i&0&0&0\\
0&0&0&i\\
0&0&-i&0
\end{pmatrix},\nonumber\\
\Gamma^{25}&=&
\begin{pmatrix}
0&0&0&i\\
0&0&i&0\\
0&-i&0&0\\
-i&0&0&0
\end{pmatrix}, \ \ 
\Gamma^{34}=
\begin{pmatrix}
0&1&0&0\\
1&0&0&0\\
0&0&0&1\\
0&0&1&0
\end{pmatrix},\nonumber\\
\Gamma^{35}&=&
\begin{pmatrix}
0&0&0&1\\
0&0&-1&0\\
0&-1&0&0\\
1&0&0&0
\end{pmatrix}, \ \ 
\Gamma^{45}=
\begin{pmatrix}
0&0&i&0\\
0&0&0&-i\\
-i&0&0&0\\
0&i&0&0
\end{pmatrix}.\nonumber\\
&&
\end{eqnarray}
$\mathcal L^{ab}$'s are ten-dimensional adjoint representation ${\bf 10}$ and 
satisfy the following SO(5) commutation relations:
\begin{eqnarray}
[\mathcal L^{ab},\mathcal L^{cd}]&=&-i(\delta_{bc}\mathcal
L^{ad}-\delta_{ac}\mathcal L^{bd}-\delta_{bd}\mathcal
L^{ac}+\delta_{ad}\mathcal L^{bc}),\ \ \ \ \ \ \ \label{f0}\\
&\equiv&if^{ab,cd,ef} {\mathcal L}^{ef},\label{f}
\end{eqnarray}
where the repeated indices are assumed to be summed over and 
$1\le a<b\le 5$, $1\le c<d\le 5$, and $1\le e<f\le 5$. $\mathcal L^{ab}$
with $a>b$ should be regarded as $\mathcal L^{ab}=-{\mathcal L}^{ba}$ on 
the right-hand side of Eq. (\ref{f0}).
\subsection{Vectors: \ {\bf 5} representations}\label{Gammamatrix}
The ${\bf 5}$ representation is an SO(5) vector and is represented by the
following five matrices: 
\begin{eqnarray}
\Gamma^{1}&=&
\begin{pmatrix}
0&0&i&0\\
0&0&0&i\\
-i&0&0&0\\
0&-i&0&0
\end{pmatrix},\ \ 
\Gamma^{2}=
\begin{pmatrix}
0&1&0&0\\
1&0&0&0\\
0&0&0&-1\\
0&0&-1&0
\end{pmatrix},\nonumber\\
\Gamma^{3}&=&
\begin{pmatrix}
0&-i&0&0\\
i&0&0&0\\
0&0&0&i\\
0&0&-i&0
\end{pmatrix},\ \  
\Gamma^{4}=
\begin{pmatrix}
1&0&0&0\\
0&-1&0&0\\
0&0&-1&0\\
0&0&0&1
\end{pmatrix},\nonumber\\
\Gamma^{5}&=&-
\begin{pmatrix}
0&0&1&0\\
0&0&0&1\\
1&0&0&0\\
0&1&0&0
\end{pmatrix}.
\end{eqnarray}
In terms of $\Gamma^a$'s, $\Gamma^{ab}$'s are represented as 
\begin{eqnarray}
\Gamma^{ab}=\frac{1}{2i}[\Gamma^a,\Gamma^b].\label{nGamma}
\end{eqnarray}
The commutation relations between $\Gamma^a$ and $\Gamma^{bc}$ are 
given by
\begin{eqnarray}
[\Gamma^a,{\Gamma}^{bc}]=-2i\Big({\rm sgn}(c-a)\delta_{ab}\Gamma^c
+{\rm sgn}(b-a)\delta_{ac}\Gamma^b\Big).\ \ \label{CommLG} 
\end{eqnarray}

\section{Axial Charge} \label{AxialC}
In this Appendix, we summarize various spherical tensors with respect to
the axial charge. They appear as a part of the SO(5) degrees of freedom
in the main text.
\subsection{Spherical tensors of axial charge symmetry}
Spherical tensor operators $T_m^{(l)}$ in the axial charge sector are defined by
\begin{eqnarray}
[{I_{\rm tot}}_z, T_m^{(l)}] &=& mT_m^{(l)},
\end{eqnarray}
\begin{eqnarray}
[{I_{\rm tot}}_{\pm}, T_m^{(l)}] &=& \sqrt{(l\mp m)(l\pm m+1)}T_{m\pm 1}^{(l)}.
\end{eqnarray}
Here, ${\bf I}_{\rm tot}$ is defined by Eqs. (\ref{Iz}) and (\ref{Ip})
and $l$ is an integer and is called the rank of operator $T^{(l)}_m$ and
$|m|\le l$. Operators with $l=0$ are scalar, {\it i.e.},  
invariant under the charge SU(2) operations, while operators with $l\ge
1$ transform as rank-$l$ tensors. In particular, operators with $l=1$
 transform as vectors. A trivial example is the axial charge of
conduction electrons ${\bf I}(x)$,
\begin{eqnarray}
I_z(x)&=&\frac{1}{2}\sum_{\sigma}\Big[
s^{\dagger}_{\sigma}(x)s_{\sigma}(x)+
p^{\dagger}_{\sigma}(x)p_{\sigma}(x)-1
\Big],\label{IzC}\\
I_+(x)&=& s_{\uparrow}^{\dagger}(x)s_{\downarrow}^{\dagger}(x)
+p_{\uparrow}^{\dagger}(x)p_{\downarrow}^{\dagger}(x)=[I_-(x)]^{\dagger}.\label{IpC}
\end{eqnarray}
${\bf I}(x)$ denotes the contributions of the conduction electrons to ${\bf
I}_{\rm tot}$\cite{axialchargeNOTE} and the rank-1 tensor with  
$[T^{(1)}_{-1},T^{(1)}_0,T^{(1)}_1]$$=$$[I_-(x)/\sqrt{2},I_z(x),-I_+(x)/\sqrt{2}]$. Thus,
the quantum number of ${\bf I}(x)$ is $(j_{j_z},i,P)=(0_0,1,0)$, where, 
$j(i)$ is the eigenvalue of spin (axial charge) with the z-component
$j_z$ and $P$ is the parity. Here, the
quantum numbers in the spin sector can be determined in the same way as
in the axial charge sector. The quantum numbers of the $f$-electron axial
charge ${\bf I}$ are the same as those of ${\bf I}(x)$.
In the following, we will list various
spherical tensors that appear in the SO(5) degrees of freedom analyzed
in the main text.
\subsection{Longitudinal-flavor axial charge}
We define longitudinal-flavor axial charge ${\bf I}'(x)$, 
which is related to the SO(5) generator 
$L^{12}(x)$, $L^{23}(x)$ and $L^{25}(x)$,
and appears in Eq. (\ref{II}) as
\begin{eqnarray}
I'_z(x)&=&\frac{1}{2}\sum_{\sigma}\Big[
s^{\dagger}_{\sigma}(x)s_{\sigma}(x)-
p^{\dagger}_{\sigma}(x)p_{\sigma}(x)
\Big],\label{I'z}\\
I'_+(x)&=& s_{\uparrow}^{\dagger}(x)s_{\downarrow}^{\dagger}(x)
-p_{\uparrow}^{\dagger}(x)p_{\downarrow}^{\dagger}(x)=[I'_-(x)]^{\dagger}.\label{I'p}
\end{eqnarray}
The parity of these operators is even,
since $p_{\sigma}$ appears as quadratic forms in Eqs. (\ref{I'z}) and (\ref{I'p}). 
In terms of the
spherical tensors, ${\bf I}'(x)$ is the rank 1 tensor 
with 
$[T^{(1)}_{-1},T^{(1)}_0,T^{(1)}_1]$$=$$[I'_-(x)/\sqrt{2},I'_z(x),-I'_+(x)/\sqrt{2}]$. 
Thus, the quantum number of ${\bf I}'(x)$ is $(j_{j_z},i,P)=(0_0,1,0)$.

\subsection{Transverse-flavor axial charge}
We define transverse-flavor axial charge ${\bf I}''(x)$, which is
related to 
$L^{14}(x)$, $L^{34}(x)$ and $L^{45}(x)$,
and 
appears in Eq. (\ref{II}) as
\begin{eqnarray}
I''_z(x)&=&\frac{1}{2}\sum_{\sigma}\Big[
p^{\dagger}_{\sigma}(x)s_{\sigma}(x)+
s^{\dagger}_{\sigma}(x)p_{\sigma}(x)
\Big],\label{I''z}\\
I''_+(x)&=& s_{\uparrow}^{\dagger}(x)p_{\downarrow}^{\dagger}(x)
+p_{\uparrow}^{\dagger}(x)s_{\downarrow}^{\dagger}(x)=[I''_-(x)]^{\dagger}.\label{I''p}
\end{eqnarray}
The parity of these operators is odd, since ${\bf I}''(x)$
includes one $p_{\sigma}$ in each term.
In terms of the
spherical tensors, ${\bf I}''(x)$ is the rank 1 tensor and
$[T^{(1)}_{-1},T^{(1)}_0,T^{(1)}_1]$$=$$[I''_-(x)/\sqrt{2},I''_z(x),-I''_+(x)/\sqrt{2}]$. 
Thus, the quantum number of ${\bf I}''(x)$ is $(j_{j_z},i,P)=(0_0,1,1)$.

\subsection{Flavor singlet}
We define a flavor singlet operator ${\mathcal D}(x)=L^{24}(x)$, which appears in Eq. (\ref{II}) as
\begin{eqnarray}
{\mathcal D}(x)&=&-\frac{i}{2}\sum_{\sigma}\Big[
s^{\dagger}_{\sigma}(x)p_{\sigma}(x)-
p^{\dagger}_{\sigma}(x)s_{\sigma}(x)
\Big].
\end{eqnarray}
The parity of this operator is odd, as is evident from the
fact that $\mathcal D(x)$ includes one $p_{\sigma}$ in each term. 
Since $[{I_{\rm tot}}_{\pm},{\mathcal D}(x)]=0$ and $[{I_{\rm
tot}}_{z},{\mathcal D}(x)]=0$, $\mathcal D(x)$ is a scalar operator and  
the quantum number of $\mathcal D(x)$ is $(j_{j_z},i,P)=(0_0,0,1)$.

\subsection{Longitudinal spin-flavor singlet}
We define a longitudinal spin-flavor singlet operator ${\mathcal
D}'(x)=n^4(x)$, {\it i.e.}, the fourth component of the SO(5) vector ${\bf n}(x)$ in Table \ref{tbl-0}  as
\begin{eqnarray}
{\mathcal D}'(x)&=&\frac{1}{2}\sum_{\sigma}\sigma\Big[
s^{\dagger}_{\sigma}(x)s_{\sigma}(x)-
p^{\dagger}_{\sigma}(x)p_{\sigma}(x)
\Big].
\end{eqnarray}
The parity of this operator is even, as is evident from the
fact that $\mathcal D'(x)$ includes zero or two $p_{\sigma}$'s in each term. 
Since $[{I_{\rm tot}}_{\pm},{\mathcal D}'(x)]=0$ and $[{I_{\rm
tot}}_{z},{\mathcal D}'(x)]=0$, $\mathcal D'(x)$ is a scalar operator and  
the quantum number of $\mathcal D'(x)$ is
$(j_{j_z},i,P)=(1_0,0,0)$.

\subsection{Transverse spin-flavor singlet}
We define a transverse spin-flavor singlet operator ${\mathcal D}''(x)=n^2(x)$  in Table \ref{tbl-0}  as
\begin{eqnarray}
{\mathcal D}''(x)&=&\frac{1}{2}\sum_{\sigma}\sigma\Big[
p^{\dagger}_{\sigma}(x)s_{\sigma}(x)+
s^{\dagger}_{\sigma}(x)p_{\sigma}(x)
\Big].
\end{eqnarray}
The parity of this operator is odd, since 
 $\mathcal D''(x)$ includes one $p_{\sigma}$ in each term. 
Since $[{I_{\rm tot}}_{\pm},{\mathcal D}''(x)]=0$ and $[{I_{\rm
tot}}_{z},{\mathcal D}''(x)]=0$, $\mathcal D''(x)$ is a scalar operator and  
the quantum number of $\mathcal D''(x)$ is $(j_{j_z},i,P)=(1_0,0,1)$.

\subsection{Transverse-spin-flavor axial charge}
We define transverse-spin-flavor axial charge ${\bf I}'''(x)$, which is
related to $n^1(x)$, $n^3(x)$ and $n^5(x)$ in Table \ref{tbl-0} as
\begin{eqnarray}
I'''_z(x)&=&\frac{i}{2}\sum_{\sigma}\sigma\Big[
p^{\dagger}_{\sigma}(x)s_{\sigma}(x)-
s^{\dagger}_{\sigma}(x)p_{\sigma}(x)
\Big],\label{I'''z}\\
I'''_+(x)&=& -i\Big[
s_{\uparrow}^{\dagger}(x)p_{\downarrow}^{\dagger}(x)
+s_{\downarrow}^{\dagger}(x)p_{\uparrow}^{\dagger}(x)
\Big]
=[I'''_-(x)]^{\dagger}.\ \ \ \ 
\ \ \ \ \label{I'''p}
\end{eqnarray}
The parity of these operators is odd, since ${\bf I}'''(x)$
includes one $p_{\sigma}$ in each term.
In terms of the
spherical tensors, ${\bf I}'''(x)$ is the rank 1 tensor with 
$[T^{(1)}_{-1},T^{(1)}_0,T^{(1)}_1]$$=$$[I'''_-(x)/\sqrt{2},I'''_z(x),-I'''_+(x)/\sqrt{2}]$. 
Thus, the quantum number of ${\bf I}'''(x)$ is $(j_{j_z},i,P)=(1_0,1,1)$.

\section{Primary states of SO(5)$_2$ Kac-Moody algebra}\label{primary}
In this Appendix, we briefly show that the primary states for $k=2$
SO(5) Kac-Moody algebra are ${\bf 1}$, ${\bf 4}$, and ${\bf 5}$
representations. 

Since the rank of SO(5) group is 2 and thus the Cartan subalgebra consists of
$H_1\equiv L_0^{15}$ and $H_2\equiv L_0^{23}$,  
we can label primary states by their eigenvalues, $h_1$ and $h_2$, and denote
them as $|h_1,h_2\rangle$ with $h_1$ and $h_2$ being integers or
half-integers. Since $|h_1,h_2\rangle$ is primary,
$L_n^{ab}|h_1,h_2\rangle=0$ for $n>0$.
Ladder operators are defined as
\begin{eqnarray}
J_-^{(1)}&\equiv& L_0^{34}+iL_0^{24}\equiv [J_+^{(1)}]^{\dagger},\\
J_-^{(2)}&\equiv& \frac{1}{2}\Big[L_0^{35}-L_0^{12}
-iL_0^{13}-iL_0^{25}
\Big]\equiv [J^{(2)}_+]^{\dagger},
\end{eqnarray}
and they satisfy
\begin{eqnarray}
&&[H_1,J_-^{(1)}]=0, \ \ \ \ \ \ \ \ \ \ [H_2,J_-^{(1)}]=-J_-^{(1)},\label{lad1}\\
&&[H_1,J_-^{(2)}]=-J_-^{(2)}, \ \ \ \ [H_2,J_-^{(2)}]=J_-^{(2)}.\label{lad2}
\end{eqnarray}
The commutation relations Eqs. (\ref{lad1}) and (\ref{lad2}) indicate 
$J_-^{(1)}|h_1,h_2\rangle\propto |h_1,h_2-1\rangle$ and 
$J_-^{(2)}|h_1,h_2\rangle\propto$$|h_1-1,h_2+1\rangle$.

Similarly, we can define another set of ladder operators as 
\begin{eqnarray}
\tilde{J}_-^{(1)}&\equiv& L_{+1}^{34}+iL_{+1}^{24}\equiv [\tilde{J}_+^{(1)}]^{\dagger},\\
\tilde{J}_-^{(2)}&\equiv& \frac{1}{2}\Big[L_{+1}^{35}-L_{+1}^{12}
-iL_{+1}^{13}-iL_{+1}^{25}
\Big]\equiv [\tilde{J}^{(2)}_+]^{\dagger},
\end{eqnarray}
where $L_{+1}^{ab}=[L_{-1}^{ab}]^{\dagger}$. 
Straightforward calculations show that they satisfy 
\begin{eqnarray}
&&[H_1,\tilde{J}_-^{(1)}]=0, \ \ \ \ \ \ \ \ \ \ [H_2,\tilde{J}_-^{(1)}]=-\tilde{J}_-^{(1)},\label{lad1tilde}\\
&&[H_1,\tilde{J}_-^{(2)}]=-\tilde{J}_-^{(2)}, \ \ \ \ [H_2,\tilde{J}_-^{(2)}]=\tilde{J}_-^{(2)},\label{lad2tilde}
\end{eqnarray}
and also satisfy
\begin{eqnarray}
[ \tilde{J}_+^{(1)}, \tilde{J}_-^{(1)} ]&=&2H_2-k,\label{JJ1}
\end{eqnarray}
and 
\begin{eqnarray}
[ \tilde{J}_+^{(2)}, \tilde{J}_-^{(2)} ]&=&H_1-H_2-k/2.\label{JJ2}
\end{eqnarray}
Now, let us consider the norm of descendant states.
Since the norm is positive, we obtain
\begin{eqnarray}
|\tilde{J}_+^{(1)}|h_1,h_2\rangle|^2&=&
\langle  h_1,h_2|\tilde{J}_-^{(1)}\tilde{J}_+^{(1)}|h_1,h_2\rangle,\nonumber\\
&=&
\langle  h_1,h_2|[\tilde{J}_-^{(1)},\tilde{J}_+^{(1)}]|h_1,h_2\rangle,\nonumber\\
&=& -2h_2+k\ge 0, \label{eqh2}
\end{eqnarray}
where at the second line we have used $L_{+1}^{ab}|h_1,h_2\rangle=0$ and
at the third line, Eq. (\ref{JJ1}) and $\langle
h_1,h_2|h_1,h_2\rangle=1$ have been used. Similar calculations  
for $\tilde{J}_+^{(2)}|h_1,h_2\rangle$ lead to
\begin{eqnarray}
h_2-h_1+k/2\ge 0. \label{eqh12}
\end{eqnarray}
It is clear that irreducible representations with the large dimension
cannot satisfy Eqs. (\ref{eqh2}) and (\ref{eqh12}), since, in general,
such irreducible representations have large $|h_1|$ and
$|h_2|$. Indeed, Eqs. (\ref{eqh2}) and (\ref{eqh12}) indicate that there
are three irreducible representations, and they are
 the identity ${\bf 1}$, the spinor ${\bf
4}$, and the vector ${\bf 5}$. Thus, primary states in the SO(5)$_2$
sector belong to ${\bf 1}$, ${\bf 4}$, or ${\bf 5}$ representations.


\begin{thebibliography}{99}
\bibitem{nonmagfix}This fixed point can be characterized by three different phenomenological 
pictures. One is to regard it as that with non-interacting
$f$-electrons plus decoupled $\vec{\tau}$. The second is to regard it as
a
system with antiferromagnetic strong coupling
spin exchange interactions between the f- and p-electrons and also 
decoupled $\vec{\tau}$. Thirdly, in our language in Sec. \ref{BCFT}, it is 
characterized by strong coupling antiferro axial charge exchange
interactions between the f- and p-electrons and again with decoupled $\vec{\tau}$.
\end{thebibliography}
\end{document}